\documentclass[pdflatex,sn-mathphys]{sn-jnl}

\usepackage{diagbox}
\usepackage{lineno}
\usepackage{natbib}

\jyear{2021}%

\theoremstyle{thmstyleone}%
\theoremstyle{thmstyletwo}%

\theoremstyle{thmstylethree}%

\begin{document}
	\title[VBQC with identity authentication for different types of clients]{Verifiable blind quantum computation with identity authentication for different types of clients}
	

	\author[1]{\fnm{Junyu Quan} }{}

	\author[2]{\fnm{Qin Li} }{}

	\author[3]{\fnm{Lvzhou Li} }

	\affil[1]{\orgdiv{School of Mathematics and Computational Science}, \orgname{Xiangtan University}, \orgaddress{ \city{Xiangtan}, \postcode{411105}, \country{China}}}
		
	\affil[2]{\orgdiv{School of Computer Science}, \orgname{Xiangtan University}, \orgaddress{ \city{Xiangtan}, \postcode{411105}, \country{China}}}
	
	\affil[3]{\orgdiv{Institute of Quantum Computing and Computer Theory,\\ School of Computer Science and
			Engineering}, \orgname{ Sun Yat-sen University}, \orgaddress{\city{Guangzhou}, \postcode{510006}, \country{China}}}
	
	
	\abstract{Quantum computing has considerable advantages in solving some problems over its classical counterpart. Currently various physical systems are developed to construct quantum computers but it is still challenging and the first use of quantum computers may adopt the cloud style. Blind quantum computing (BQC) provides a solution for clients with limited quantum capabilities to delegate their quantum computation to remote quantum servers while keeping input, output, and even algorithm private. In this paper, we propose three multi-party verifiable blind quantum computing (VBQC) protocols with identity authentication to handle clients with varying quantum capabilities in quantum networks, such as those who can just make measurements, prepare single qubits, or perform a few single-qubit gates. They are client-friendly and flexible since the clients can achieve BQC depending on their own quantum devices and resist both insider outsider attacks in quantum networks. Furthermore, all the three proposed protocols are verifiable, namely that the clients can verify the correctness of their calculations.}

	
	

\maketitle

\section{Introduction}\label{sec1}

Quantum computation is a revolutionary computer model that has the potential to provide significant computational benefits over traditional technologies by utilizing principles of quantum mechanics. For instance, Shor's algorithm can solve the integer factorization problem in polynomial time \cite{shor1994algorithms}, whereas the best existing classical algorithm requires super-polynomial time. Even though quantum computing has several benefits and numerous physical systems such as trapped ions \cite{leibfried2003quantum,blatt2012quantum}, superconducting qubits \cite{krantz2019a,kjaergaard2020superconducting}, photons \cite{XiLinWang201818QubitEW,wang2019boson}, and silicon \cite{YuHe2019ATG,hensen2020a} have been investigated for building quantum computers, it remains a challenge to construct large-scale quantum computers. In the near future, it is likely that only a few organizations will own quantum computers and other clients with limited quantum capabilities have to delegate their quantum computational tasks to quantum servers. Blind quantum computing (BQC) could make clients interact securely with a remote quantum server, that is, BQC enables a client with insufficient quantum capability to delegate her quantum computation to a remote quantum server while keeping input, output, and even algorithm private.

Childs presented the first BQC protocol based on the circuit model in 2005 \cite{childs2005secure}, where the client requires quantum memory and the ability to rearrange qubits and perform SWAP gates. Then, Arrighi and Salvail introduced another BQC protocol that only requires Alice to generate and measure entangled states \cite{arrighi2006blind}. But it only provides both verification and blindness for certain special functions. In 2009, Broadbent, Fitzsimons, and Kashefi presented the first universal BQC protocol based on the measurement model where the client just prepares random single qubits \cite{broadbent2009universal}, and it is called the BFK protocol. Furthermore, through the use of four-qubit blind cluster states, the BFK protocol has been physically realized by Barz et al. \cite{barz2012demonstration}. In addition, the single-server BFK protocol is expanded to double-server \cite{morimae2013secure,sheng2015deterministic}, triple-server \cite{li2014triple}, and multiple-server modes \cite{kong2016multiple}. In the double-server BQC protocols proposed in Refs. \cite{morimae2013secure,sheng2015deterministic}, the client can be completely classical if the two delegated quantum servers are supposed to be non-communicating. Then Li et al. proposed a triple-server BQC protocol where communication is allowed among servers via entanglement swapping \cite{li2014triple}. Different from the BFK protocol, Morimae and Fujii proposed a BQC protocol called the MF protocol based on a new model \cite{morimae2013blind} in which the client only makes measurements. This protocol has also been demonstrated by using four-qubit cluster states \cite{greganti2016demonstration}. Since then, numerous BQC protocols based on various methods were developed \cite{reichardt2013classical,huang2017experimental,2015Iterated,xu2022universal,li2021blind,li2021quantum}.

Almost all the proposed BQC protocols should satisfy two important properties, namely blindness and correctness. Blindness prevents the delegated quantum server from obtaining any valuable information about the client's input, output, and algorithm, while correctness means the client could verify whether the computation results are right, which can be ensured by verifiable blind quantum computation (VBQC) \cite{fitzsimons2017unconditionally,fitzsimons2018post,hajdusek2015device,morimae2017verification,takeuchi2018verification}. Currently, VBQC protocols are mainly divided into two types, namely stabilizer test verification \cite{hayashi2015verifiable,morimae2017verification,takeuchi2018verification} and trap-based verification \cite{fitzsimons2017unconditionally}. For the first type, the server generates multiple copies of graph states and sends each qubit of them one by one to the client. Then the client verifies the stabilizer operators of these graph states to determine whether the server has constructed correct graph states. Trap-based verification, on the other hand, requires a client generating single-qubit states and embedding trap qubits to check the server's behaviour. If the server does not follow the protocol, he may disturb the trap qubits which could be detected by the client with a high probability. 

In addition, we are moving into the noisy intermediate-scale quantum (NISQ) era now \cite{preskill2019quantum}, which means one can control a quantum system with qubits from 50 to a few hundred \cite{zhang2017observation,arute2019quantum}. More quantum nodes are expected to arise in quantum networks as a result of the development of quantum computing \cite{GuanYuWang2020EntanglementPF,ZhiHaoLiu221,WenJunShi1291}. In such a complex network environment, the classical clients who want to perform BQC will face more problems compared with that in a single environment. For example, in a multi-node quantum network, outsider attacks from malicious third parties during transmission and effective authentication between individual nodes need to be considered; due to the decoherence caused by noise, clients need to verify the correctness of their calculation results even if the server is honest; and how different clients with various quantum devices to perform BQC securely also need consideration.

Recently, Li et al. first introduced identity authentication to BQC for resisting the man-in-the-middle attack and denial-of-service attack \cite{li2018blind}. Then Shan et al. presented a BQC protocol with mutual authentication based on quantum network \cite{shan2021multi}, which includes registration phase, mutual identity authentication phase, and multi-party blind quantum computation phase. In this protocol, there are numerous clients and servers. Each legitimate client needs to establish a registration key with a specific server through certificate authority (CA) and load balancers in the registration phase. After the client and the corresponding server authenticate each other with the registration key, she can delegate her quantum computation to the server. However, only the clients who are limited to the capability of making measurements are considered in this protocol. In fact, clients in various BQC protocols have different quantum capabilities and are suitable for special physical systems. In an optical system, for instance, measuring a single-qubit state is easier than generating a single qubit \cite{morimae2013blind}, hence the MF protocol is more favorable to clients who possess measurement devices in this situation. In superconducting systems, performing single-qubit gates is more precise than preparing or measuring single-qubit states \cite{barends2014superconducting}. Hence the protocol in which clients only perform a few quantum gates \cite{li2021blind} is better for superconducting system. Therefore, it is necessary to consider how to enable clients to delegate their quantum computation securely depending on their different quantum capabilities in a quantum network. Furthermore, in the BQC protocol proposed in Ref. \cite{shan2021multi}, the registration key shared by the client and the corresponding server could be obtained by the load balancers, and it is unrealistic that both the servers and the load balancers are expected to be trustworthy. Moreover, the clients cannot verify the correctness of their computation. 

Considering all the above, in this paper we propose three VBQC protocols with identity authentication for clients with different capabilities in the quantum network as shown in Table \ref{tab1.1}. Especially, in the proposed protocols, even if CA, load balancers, and servers are regarded as semi-honest, the clients can verify their computation successfully.

\begin{table*}[!htb]
	\centering
	\label{tab1.1}
	\caption{ Three Proposed Multi-party VBQC Protocols for Different Types of Clients}
	{\footnotesize
		\begin{tabular*}{\linewidth}{p{60pt}p{60pt}p{60pt}p{38pt}p{55pt}}\hline\hline
			
			Protocols & Participants  & The quantum capability of clients & Verifiability  & Identity authentication \\\hline
			
			RAM-SMVBQC in Sec. \ref{Sec4.1} &  Semi-honest CA, load balancers, and servers & Measuring single-qubit states  & Yes  & Yes \\
			
			PAS-SMVBQC in Sec. \ref{Sec4.2}  & Semi-honest CA, load balancers, and servers & Preparing single-qubit states  & Yes & Yes\\
			
			CB-SMVBQC in Sec. \ref{Sec4.3} & Semi-honest CA, load balancers, and servers & Implementing $H$ and $\sigma_z^{1/4}$ gates & Yes & Yes 
			\\\hline\hline
		\end{tabular*}
	}
\end{table*}

The rest part of the paper is organized as follows. Section \ref{Sec2} introduces two typical verification methods in VBQC protocols. Section \ref{Sec3} briefly reviews the BQC protocol proposed by Shan et al. \cite{shan2021multi}. Section \ref{Sec4} proposes three VBQC protocols. Section \ref{Sec5} compares the proposed protocols with the similar ones in the literature. The last section \ref{Sec6} makes a conclusion.

\section{Preliminaries}\label{Sec2}
In this part, the stabilizer test and trap-based verification which are two important verification methods are introduced.
\subsection{Stabilizer test}
Standard graph states and related methods for the stabilizer test \cite{hayashi2015verifiable,morimae2017verification,takeuchi2018verification} are introduced briefly as follows. An $n$-qubit graph state
$G\equiv(V,E)$ corresponding to graph $ G $ is defined as
\begin{equation}
	\label{E1}
	\vert G \rangle\equiv(\underset{e\in E}{\bigotimes}CZ_e)\vert+\rangle^{\otimes n}
\end{equation}
where $ V $ is a set of $n$ vertices, $ E $ is a set of edges, $ n\equiv\vert V\vert $, $ e\in E $, $ \vert+\rangle\equiv\frac{1}{\sqrt{2}}(\vert0\rangle+\vert1\rangle) $, and $ CZ_e $ is the Controlled-$Z$ gate ($ CZ\equiv\vert0\rangle\langle0\vert\bigotimes I +\vert1\rangle\langle1\vert\bigotimes Z $) acting on the pair of vertices sharing the edge $ e $. Then the description of $ i $th stabilizer $ g_i $ ($ 1\leq i \leq n $) of $ \vert G\rangle $ is given as follows

\begin{equation}
	\begin{split}
		g_i &\equiv(\underset{e \in E}{\prod}CZ_e)X_i(\underset{e \in E}{\prod}CZ_e ^\dagger) \\&=X_i \underset{V_j \in N^{(i)}}{\prod}Z_j
	\end{split} 	
\end{equation}
where $ N^{(i)} $ is the set of neighbors of the $ i $th qubit. About the stabilizer test for $ g_i $ on an $ n $-qubit quantum state $ \rho $, Alice measures $ X_i $ and $ Z_j $ for all $ j \in N^{(i)} $ and the measurement results are recorded as $ x_i \in \{0,1\} $ and $ z_j \in \{0,1\} $. If
\begin{equation}
	x_i + \underset{j \in N^{(i)}}{\sum z_j} \equiv 0\mod 2,
\end{equation}
Alice will pass the stabilizer test, which means that the state $\rho$ is close to the ideal state $\vert G\rangle$. Otherwise, Alice rejects it as $\rho$ is far from $\vert G\rangle$. Obviously, we can get the acceptance probability satisfies $(1+\langle G\vert\rho\vert G\rangle)/2$. If Bob is honest, he will generate the correct graph state which always satisfies Eq. (\ref{E1}) and Alice will pass the stabilizer test for $ g_i $ with unit probability for all $ i $.

\subsection{Trap-based verification}
The trap-based verification was introduced in the verifiable universal blind quantum computing (VUBQC) protocol by Fitzsimons and Kashefi \cite{fitzsimons2017unconditionally}. This protocol can achieve blindness with the method in UBQC \cite{broadbent2009universal} and accomplish verifiability through embedding trap qubits.

In UBQC, Bob produces a universal graph state $\vert G\rangle$ by using $n$ qubits $\vert+_{\theta_i}\rangle=\frac{1}{\sqrt{2}}(\vert0\rangle+e^{i\theta_i}\vert1\rangle) (i=1,2,...,n)$ sent by Alice, where $\theta_i$ is chosen from the set $\{0,\pi/4,2\pi/4,...,7\pi/4\}$. Then, Alice requests Bob to measure the $i$-th qubit of $\vert G\rangle$ in the basis $\{\vert\pm_{\delta_i}\rangle \}$, where $\delta_i=\phi_i^{'}+\theta_i + r_i\pi$, $\phi_i^{'}$ is determined by previous measurements and $ r_i$ is chosen from the set $\{0,1\}$ at random. Note that, Bob also sends the measurement result $b_i$ to Alice, which will be used to update the angles related to computation. Then Alice will obtain the output after Bob has completed all measurements. To verify the computation, Alice can prepare qubits in the state $\vert z\rangle$, which are randomly selected from $\{\vert 0\rangle, \vert 1\rangle\}$ and called dummy qubits. According to Eq. (\ref{eq:1}) and Eq. (\ref{eq:2}),
\begin{equation}
	\begin{aligned}
		\label{eq:1}
		CZ(\vert0\rangle \otimes\vert+_{\theta}\rangle) &= \vert0\rangle\otimes\vert0\rangle+\vert0\rangle\otimes e^{i\theta}\vert1\rangle
		\\	&=\vert0\rangle\otimes\vert+_{\theta}\rangle,
	\end{aligned}
\end{equation}
\begin{equation}
	\begin{aligned}
		\label{eq:2}
		CZ(\vert1\rangle \otimes\vert+_{\theta}\rangle) &= \vert1\rangle\otimes\vert0\rangle-\vert1\rangle\otimes e^{i\theta}\vert1\rangle
		\\	&=\vert1\rangle\otimes\vert-_{\theta}\rangle,
	\end{aligned} 	
\end{equation}
a dummy qubit $\vert0\rangle$ or $\vert1\rangle$ will not entangle with the rest of the qubits $\vert+_{\theta_i}\rangle$ of the graph state and thus it will not influence the correctness or blindness of the computation. Suppose that Alice owns a trap qubit $\vert+_\theta\rangle$, she may employ some dummy qubits to isolate $\vert+_\theta\rangle$ from other adjacent states used for computation. If the trap qubit is measured in the specified basis, then the measurement result is also certain. However, from Eq. (\ref{eq:2}), if a $CZ$ gate is applied to a dummy qubit $\vert1\rangle$ and a trap qubit $\vert+_{\theta}\rangle$, the trap qubit will flip to $\vert-_{\theta}\rangle$, where the measurement result will be flipped as well. Thus, Alice can check Bob's measurements and the entangling operations to verify the results by embedding the trap qubits and the dummy qubits.

\section{Review of Shan et al.'s BQC protocol with mutual authentication \cite{shan2021multi}}
\label{Sec3}

An overview of a multi-party BQC protocol with mutual authentication proposed by Shan et al. is given as follows \cite{shan2021multi}. It involves $m$ clients $A_{i}$ ($1\leq i\leq m$), $n$ servers $B_j$ ($1\leq j\leq n$), Load\_Balancer\_A and Load\_Balancer\_B who allocates resources to clients and servers respectively, and a semi-honest certificate authority (CA). Note that the number of servers may be smaller than that of clients and then the server associated with client $A_{i}$ should be denoted by $B_{i\ mod\ n}$. The entire protocol consists of three phases, namely the registration phase mainly for quantum key distribution, the mutual identity authentication phase, and the blind quantum computing phase. In the following, the specific steps of the protocol are given.

Phase 1: the registration phase for quantum key distribution

S1-1 Load\_Balancer\_A and Load\_Balancer\_B generate qubit sequences $S_A$ and $S_B$ randomly in X or Z basis and decoy qubits in $\{\vert0\rangle,\vert1\rangle,\vert+\rangle,\vert-\rangle\}$ with suitable number. Then the decoy qubits are randomly inserted into $S_A$ and $S_B$. Load\_Balancer\_A and Load\_Balancer\_B transmit the new sequences $S_{A\_decoy}$ and $S_{B\_decoy}$ to the client $A_i$ and the corresponding server $B_{i\ mod\ n}$.

S1-2 Once $A_i$ and $B_{i\ mod\ n}$ received $S_{A\_decoy}$ and $S_{B\_decoy}$, respectively, they need to measure the decoy qubits. The positions and corresponding bases of them are informed by Load\_Balancer\_A and Load\_Balancer\_B. If the error rate is higher than the predefined threshold, an eavesdropper Eve is considered to exist and the protocol should restart again. Otherwise, Load\_Balancer\_A and Load\_Balancer\_B send $S_A$ and $S_B$ to CA.

S1-3 CA performs Bell measurements and records the measurement results as $R_{AB}$ when he receives both sequences $S_A$ and $S_B$. Note that the measurement results $\vert\phi^+\rangle$, $\vert\phi^-\rangle$, $\vert\psi^+\rangle $ and $\vert\psi^-\rangle $ are encoded as 00, 01, 10, 11, respectively. Then CA sends $R_{AB}$ to $A_i$ and $B_{i\ mod\ n}$ through Load\_Balancer\_A and Load\_Balancer\_B across classical channels.

S1-4 When $A_i$ and $B_{i\ mod\ n}$ receive $R_{AB}$, Load\_Balancer\_A and Load\_Balancer\_B announces the basis of each state in $S_A$ and $S_B$. According to Table \ref{tab2}, they keep the $j$-th bit under the same basis as the initial raw key bit $K_{AB}^j \in \{0,1\}$ when $R_{AB}^j=11$.

S1-5 $B_{i\ mod\ n}$ and $A_i$ both select a portion of their raw keys to estimate the error rate and detect eavesdropping. If the error rate is higher than the threshold, $A_i$ has to terminate the protocol and move to S1-1.

S1-6 The procedures above are repeated until each client has completed the registration and shared a key with a specific network server. If $m>n$ and $i>n$, $B_{i\ mod\ n}$ may store more than one key in his memories.

\vspace{1mm}
\tabcolsep 9pt
\renewcommand\arraystretch{1.3}
\begin{table*}[!htb]
	\centering
	\caption{\label{tab2} Correlation of Non-orthogonal States Produced by CA and Measurement Results on $S^{'}_A$ and $S^{'}_B$. }
	{\footnotesize
		
		\doublerulesep 0.1pt \tabcolsep 12pt 
		\begin{tabular*}{\linewidth}{c|cccc}\hline 
			\diagbox{$R_{A_i}$}{$\vert\phi\rangle_{AB}$}{$R_{B_{i\ mod\ n}}$}& $\vert0\rangle_{B}\rightarrow0$  & $\vert1\rangle_{B}\rightarrow1$ & $\vert+\rangle_{B}\rightarrow0$ & $\vert-\rangle_{B}\rightarrow1$\\\hline
			$\vert0\rangle_{A}\rightarrow0$ & $\vert\phi^{-}\rangle_{AB}$ & $\vert\psi^{+}\rangle_{AB}$ & $\vert\Psi^{+}\rangle_{AB}$ & $\vert\Phi^{-}\rangle_{AB}$\\
			$\vert1\rangle_{A}\rightarrow1$ & $\vert\psi^{+}\rangle_{AB}$ & $\vert\phi^{-}\rangle_{AB}$ & $\vert\Phi^{-}\rangle_{AB}$ & $\vert\Psi^{+}\rangle_{AB}$\\
			$\vert+\rangle_{A}\rightarrow0$ & $\vert\Psi^{+}\rangle_{AB}$ & $\vert\Phi^{-}\rangle_{AB}$ & $\vert\psi^{+}\rangle_{AB}$ & $\vert\phi^{-}\rangle_{AB}$\\
			$\vert-\rangle_{A}\rightarrow1$ & $\vert\Phi^{-}\rangle_{AB}$ & $\vert\Psi^{+}\rangle_{AB}$ & $\vert\phi^{-}\rangle_{AB}$ & $\vert\psi^{+}\rangle_{AB}$
			\\\hline
		\end{tabular*}
	}
\end{table*}
\baselineskip=18pt plus.2pt minus.2pt
\parskip=0pt plus.2pt minus0.2pt

Phase 2: the mutual identity authentication phase

S2-1 If a registered client $A_i$ wants to delegate a BQC task to a remote server, he needs to send a request $i$ to Load\_Balancer\_A firstly.

S2-2 According to the first-in-first-out (FIFO) principle, Load\_Balancer\_A resends the request $i$ to CA, who prepares $4k$ non-orthogonal states $\vert\varphi\rangle_{AB}$ randomly chosen from $\{\vert\phi^{-}\rangle,\vert\psi^{+}\rangle,\vert\Phi^{-}\rangle,\vert\Psi^{+}\rangle\}$, where
\begin{equation}
	\label{eq7}
	\begin{split}
		\vert\phi^{-}\rangle_{AB}=&\frac{1}{\sqrt{2}}(\vert00\rangle-\vert11\rangle)_{AB},\\
		\vert\psi^{+}\rangle_{AB}=&\frac{1}{\sqrt{2}}(\vert01\rangle+\vert10\rangle)_{AB},\\
		\vert\Phi^{-}\rangle_{AB}=&\frac{1}{\sqrt{2}}(\vert\phi^{-}\rangle-\vert\psi^{+}\rangle)_{AB}\\
		=&\frac{1}{\sqrt{2}}(\vert0-\rangle-\vert1+\rangle)_{AB}=\frac{1}{\sqrt{2}}(\vert+1\rangle-\vert-0\rangle)_{AB},\\
		\vert\Psi^{+}\rangle_{AB}=&\frac{1}{\sqrt{2}}(\vert\phi^{-}\rangle+\vert\psi^{+}\rangle)_{AB}\\
		=&\frac{1}{\sqrt{2}}(\vert0+\rangle-\vert1-\rangle)_{AB}=\frac{1}{\sqrt{2}}(\vert+0\rangle-\vert-1\rangle)_{AB}.
	\end{split}
\end{equation}
All the first qubits of these non-orthogonal states form the qubit sequence $S^{'}_A$ and the second qubits of them form the qubit sequence $S^{'}_B$. CA sends $S^{'}_A$ and $S^{'}_B$ to Load\_Balancer\_A and Load\_Balancer\_B, respectively. Note that, CA also encodes the non-orthogonal states as $\vert\phi^{-}\rangle \to 00$, $\vert\psi^{+}\rangle \to 01$, $\vert\Phi^{-}\rangle \to 10$, $\vert\Psi^{+}\rangle \to 11$, and broadcasts them to $A_i$ and $B_{i\ mod\ n}$. When Load\_Balancer\_A and Load\_Balancer\_B receive $S^{'}_A$ and $S^{'}_B$, they generate some decoy qubits and insert them into $S^{'}_A$ and $S^{'}_B$. Then they send the new qubit sequences $S^{'}_{A\_decoy}$ and $S^{'}_{B\_decoy}$ to $A_i$ and $B_{i\ mod\ n}$, respectively.

S2-3 $A_i$ and $B_{i\ mod\ n}$ perform eavesdropping detection by checking whether the states of decoy qubits are changed. If the error rate is over a certain threshold, the step S2-2 should be re-executed; otherwise they can obtain $S^{'}_A$ and $S^{'}_B$.

S2-4 $A_i$ and $B_{i\ mod\ n}$ measure $S^{'}_A$ and $S^{'}_B$ in the bases in terms of the initial key $K_{AB}$ and record the measurement results $R_{A_i}$ and $R_{B_{i\ mod\ n}}$. Then $A_i$ and $B_{i\ mod\ n}$ randomly reveal $3k$ values of $R_{A_i}$ and $R_{B_{i\ mod\ n}}$ and the corresponding positions via classical channels.

S2-5 According to Table \ref{tab:3}, $A_i$ and $B_{i\ mod\ n}$ check independently whether the other side's measurement results are correct and complete the mutual identity authentication. For example, if CA prepares a non-orthogonal states $\vert\phi^{-}\rangle_{AB}$ and the measurement result $R_{A_i}=1$, then $R_{B_{i\ mod\ n}}=1$.

\begin{table*}[!htb]
	\tabcolsep 9pt
	\centering
	\caption{The Probabilities of CA's Bell Basis Measurement Results according to $S_A$ and $S_B$.}\vspace{-2mm}
	{\footnotesize
		\doublerulesep 0.1pt \tabcolsep 8pt 
		\begin{tabular*}{\linewidth}{c|cccc}\hline
			\label{tab:3}
			\diagbox{$S_A,S_B$}{Probabilities}{$R_{AB}$}& $\vert\phi^+\rangle_{AB}$$\rightarrow 00$ & $\vert\phi^-\rangle_{AB}\rightarrow 01$ & $\vert\psi^+\rangle_{AB}\rightarrow 10$ & $\vert\psi^-\rangle_{AB}\rightarrow 11$\\\hline
			$\vert0\rangle_{A}\vert0\rangle_{B}$ & 1/2 & 1/2 & 0 & 0\\
			$\vert0\rangle_{A}\vert1\rangle_{B}$ & 0 & 0 & 1/2 & 1/2\\
			$\vert1\rangle_{A}\vert0\rangle_{B}$ & 0 & 0 & 1/2 & 1/2\\
			$\vert1\rangle_{A}\vert1\rangle_{B}$ & 1/2 & 1/2 & 0 & 0\\
			$\vert+\rangle_{A}\vert+\rangle_{B}$ & 1/2 & 0 & 1/2 & 0\\
			$\vert+\rangle_{A}\vert-\rangle_{B}$ & 0 & 1/2 & 0 & 1/2\\
			$\vert-\rangle_{A}\vert+\rangle_{B}$ & 0 & 1/2 & 0 & 1/2\\
			$\vert-\rangle_{A}\vert-\rangle_{B}$ & 1/2 & 0 & 1/2 & 0
			\\\hline
		\end{tabular*}
	}
\end{table*}

Phase 3: the blind quantum computation phase

S3-1 After $A_i$ and $B_{i\ mod\ n}$ authenticate each other successfully, $A_i$ needs to ask $B_{i\ mod\ n}$ to generate a quantum resource state $\vert G\rangle$. Note that, a qubit of the resource state is $\vert G_{x,y}\rangle$, where $x$ denotes a row number and $y$ denotes a column number.

S3-2 $B_{i\ mod\ n}$ sends these qubits of the resource state to $A_i$ via quantum channels one by one.

S3-3 $A_i$ measures these qubits in basis $\{\vert\pm_{\theta^{'}_{x,y}}\rangle\}$, where
$\theta^{'}_{x,y}=(-1)^{s_{x,y}^X}\theta_{x,y}+s_{x,y}^Z\pi$. Note that $\theta_{x,y}\in\{-\frac{\pi}{4},0,\frac{\pi}{4},\frac{\pi}{2}\}$ is the desired measurement angle, $s_{x,y}^X$ is the summation of all previous measurement results in $X$ basis, $s_{x,y}^Z$ is the summation of all previous measurement results in $Z$ basis, and the measurement result is recorded as $s_{x,y}\in\{0,1\}$. If all qubits have been measured, the computation is completed.

As mentioned above, the client shares a key $K_{AB}$ with the specific server under the help of load balancers and CA in the registration phase. Then, if the client wants to perform a BQC task, she can log into the remote server by using the key $K_{AB}$ in the identity authentication phase. After the client and the server authenticate each other, the client can delegate her BQC task to the server securely. Here a brief analysis of the security of the reviewed protocol is given. Firstly, a reasonable assumption is that an unjammable public channel is required to ensure the integrity of transmitted classical messages between each load balancer and CA. Therefore, in Phase 1 and Phase 2, as long as the devices for preparing single-qubit states owned by Load\_Balancer\_A and Load\_Balancer\_B are secure, neither the semi-honest CA nor an external attacker Eve can obtain the shared key $K_{AB}$. In Phase 3, based on the no-signaling principle, the privacy of the clients can be well protected. In addition, because the decoy-state check technology is used in each quantum information transmission, this protocol can resist intercept-resend attack and entangle-measure attack. However, Load\_Balancer\_A and Load\_Balancer\_B are regarded as honest and they can get information about $S_A$, $S_B$, $R_{AB}$, from which the value of the shared key can be deduced. Once the load balancers are successfully attacked, the entire network is no longer secure. What's more, in this protocol, only the clients who can make measurements are considered. Actually, the clients usually have different quantum capabilities in a real quantum network. Especially, the clients cannot determine whether the nodes in the quantum network are honest or not. Therefore, it is necessary to consider the verifiability of their quantum computation.

\section{Three proposed multi-party VBQC protocols}
\label{Sec4}

Since the clients in quantum networks have different quantum capabilities, such as making measurements, generating single-qubit states, or performing a few single-qubit gates, we propose three multi-party VBQC protocols with identity authentication suitable for three types of clients based on the mainly blind quantum computing model. 

\subsection{The first secure multi-party VBQC protocol based on the receive-and-measure method}
\label{Sec4.1}
In this part, we construct a secure multi-party VBQC protocol mainly based on the receive-and-measure method (called RAM-SMVBQC). Similar to Shan et al.'s BQC protocol \cite{shan2021multi}, in this protocol, the clients only have the ability to make measurements and own a little quantum memory. First, the clients should share a register key $K_{AB}$ with the specific server with the help of load balancers and CA. Then, the clients could employ the key $K_{AB}$ to access the remote quantum server to delegate their quantum computational tasks. The specific steps are as follows.

Phase 1: the registration phase

A1-1 $A_i$ first sends a request $i$ to Load\_Balancer\_A who puts it into the request queue. Conforming to the FIFO principle, Load\_Balancer\_A forwards the request $i$ to Load\_Balancer\_B and CA. Then Load\_Balancer\_A generates a series of qubits randomly in $\{\vert0\rangle,\vert1\rangle,\vert+\rangle,\vert-\rangle\}$ with the length of $m+l+k$. The qubit sequence is denoted by $S_{LA_{decoy}}$ and is sent to $A_i$.

A1-2 When $A_i$ receives $S_{LA_{decoy}}$, she randomly measures $l$ qubits in the corresponding bases, which should be told by Load\_Balancer\_A, and returns $S_{A_{decoy}}$ consisting of the rest of $m+k$ qubits to Load\_Balancer\_A. If the measurement results are correct, $A_i$ tells Load\_Balancer\_A the positions and bases of other $m$ qubits. Load\_Balancer\_A measures them in the informed measurement bases. The remaining qubit sequence is represented by $S_A$. Either of the error rates in these two decoy detections is higher than the predefined threshold, $A_i$ has to terminate the protocol and move to A1-1. Note that if $A_i$ owns massive quantum memory, she can do it efficiently. Otherwise, she needs to set the clock period carefully to ensure that Bob cannot distinguish which qubits she measured.

A1-3 After Load\_Balancer\_B receives the request $i$, he assigns an idle server $B_{i\ mod\ n}$ and transmits the request $i$ to $B_{i\ mod\ n}$. Then $B_{i\ mod\ n}$ randomly generates $m+k$ qubits in $\{\vert0\rangle,\vert1\rangle,\vert+\rangle,\vert-\rangle\}$, which form the sequence $S_{B_{decoy}}$. Next he sends $S_{B_{decoy}}$ to Load\_Balancer\_B. Similar to A1-2, after Load\_Balancer\_B receives these qubits, he has to perform the decoy detection and the remaining qubit sequence is represented by $S_B$.

A1-4 CA performs Bell measurements on $S_A$ and $S_B$, which are sent from Load\_Balancer\_A and Load\_Balancer\_B, and records the measurement results as $R_{AB}$. The measurement results $\{\vert\phi^+\rangle ,\vert\psi^+\rangle, \vert\phi^-\rangle, \vert\psi^-\rangle\}$ are recorded as $\{00, 01, 10, 11\}$, respectively. Then, CA sends $R_{AB}$ to Load\_Balancer\_A and Load\_Balancer\_B via public classical channels. Finally, they resend $R_{AB}$ to $A_i$ and $B_{i\ mod\ n}$.

A1-5 When $A_i$ and $B_{i\ mod\ n}$ receive $R_{AB}$ successfully, they have to securely exchange the bases of qubits in $S_A$ and $S_B$. Similarly, according to Table \ref{tab2}, they keep the $j$-th bit under the same basis as the initial raw key bit $K_{AB}^j \in \{0,1\}$ when $R_{AB}^j=11$. For example, if the $j$-th qubit in $S_A$ or $S_B$ is prepared in $Z$ basis, then $K_{AB}^j=0$. Otherwise, $K_{AB}^j=1$.

A1-6 $A_i$ and $B_{i\ mod\ n}$ have to pick a portion of their raw keys to estimate error rate and detect eavesdropping as a consequence of the effect of noise in the actual channel. If the error rate is below the threshold, the channel is secure. Otherwise, $A_i$ has to terminate protocol and move to A1-1.

\begin{figure*}[!t]
	\centering
	\includegraphics[width=0.8\textwidth]{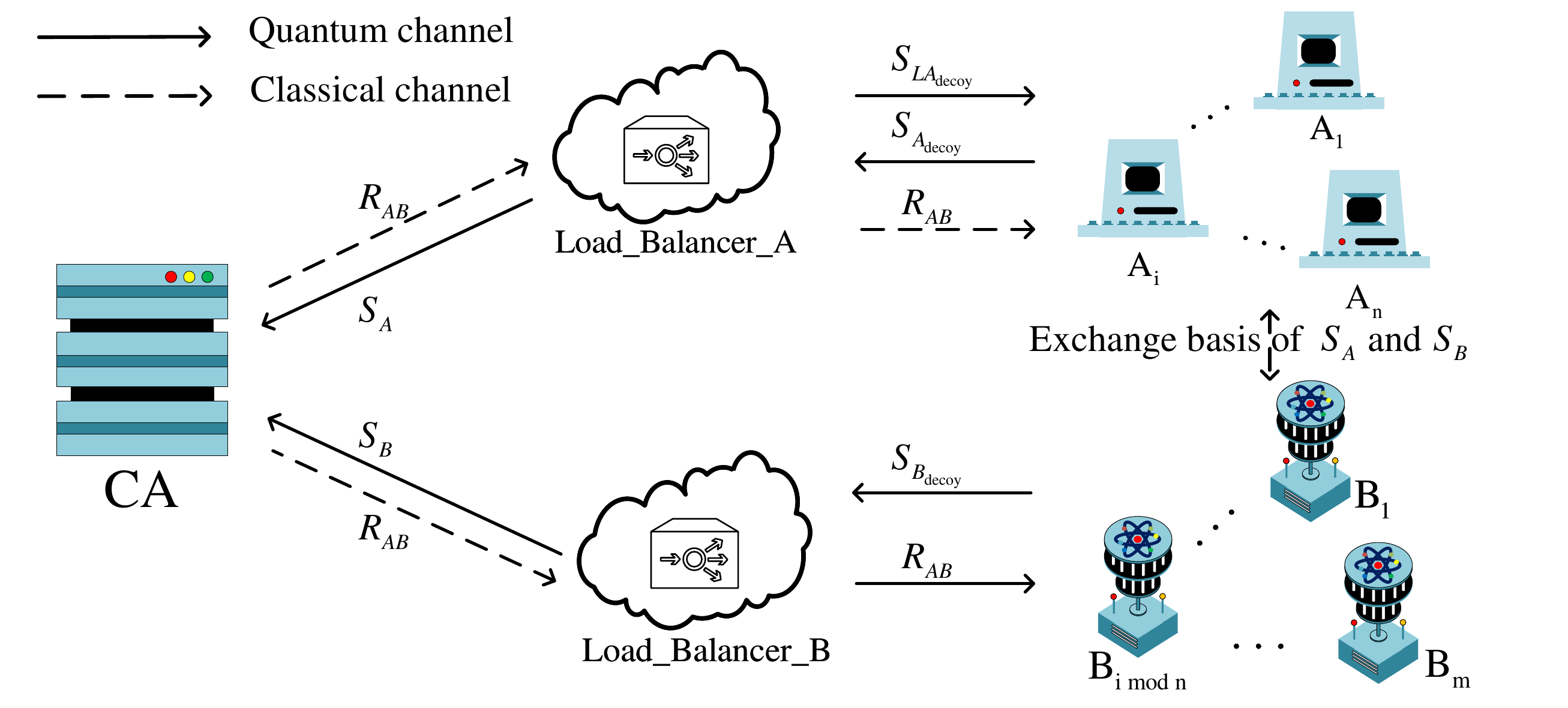}
	\caption{(Color Online) Information flow of the registration phase in the RAM-SMVBQC protocol}
	\label{fig1}
\end{figure*}

A1-7 The steps from A1-1 to A1-6 should be repeated until all clients have completed the registration and each of them has obtained a shared key with a specific network server, as illustrated in Fig. \ref{fig1}. Note that, if $m>n$ and $i>n$, $B_{i\ mod\ n}$ must securely store multiple keys in his memories.

Phase 2: the identity authentication phase

A2-1 $A_i$ sends request $i$ to Load\_Balancer\_A when he wants to delegate a BQC to a remote server.

A2-2 Based on the FIFO principle, Load\_Balancer\_A resends the request $i$ to CA, who needs to randomly prepare $4k$ non-orthogonal states selected in $\{ \vert\phi^{-}\rangle,\vert\psi^{+}\rangle,\vert\Phi^{-}\rangle,\vert\Psi^{+}\rangle \}$ as shown in Eq. (\ref{eq7}). All the first qubits of $\vert\varphi\rangle_{AB}$ form the qubit sequence $S^{'}_A$ which should be sent to Load\_Balancer\_A and the second qubits of them form the qubit sequence $S^{'}_B$ which should be sent to Load\_Balancer\_B. Note that, CA also encodes the non-orthogonal states as $S_{AB}$ which denoted $\vert\phi^+\rangle \to 00$, $\vert\phi^-\rangle \to 01$, $\vert\psi^+\rangle \to 01$, $\vert\psi^-\rangle \to 11$, and broadcasts them to $A_i$ and $B_{i\ mod\ n}$. When Load\_Balancer\_A and Load\_Balancer\_B receive $S^{'}_A$ and $S^{'}_B$, they insert some decoy qubits into $S^{'}_A$ and $S^{'}_B$ and send the new qubit sequences $S^{'}_{A_{decoy}}$ and $S^{'}_{B_{decoy}}$ to $A_i$ and $B_{i\ mod\ n}$, respectively.

A2-3 After eavesdropping detection, $A_i$ and $B_{i\ mod\ n}$ measure $S^{'}_A$ and $S^{'}_B$ in the bases in terms of the initial key $K_{AB}$ and record the measurement results $R_{A_i}$ and $R_{B_{i\ mod\ n}}$. Then $A_i$ and $B_{i\ mod\ n}$ randomly reveal $3k$ values of $R_{A_i}$ and $R_{B_{i\ mod\ n}}$ and the corresponding positions via classical channels.

\begin{figure*}[!t]
	\centering
	\includegraphics[width=0.8\textwidth]{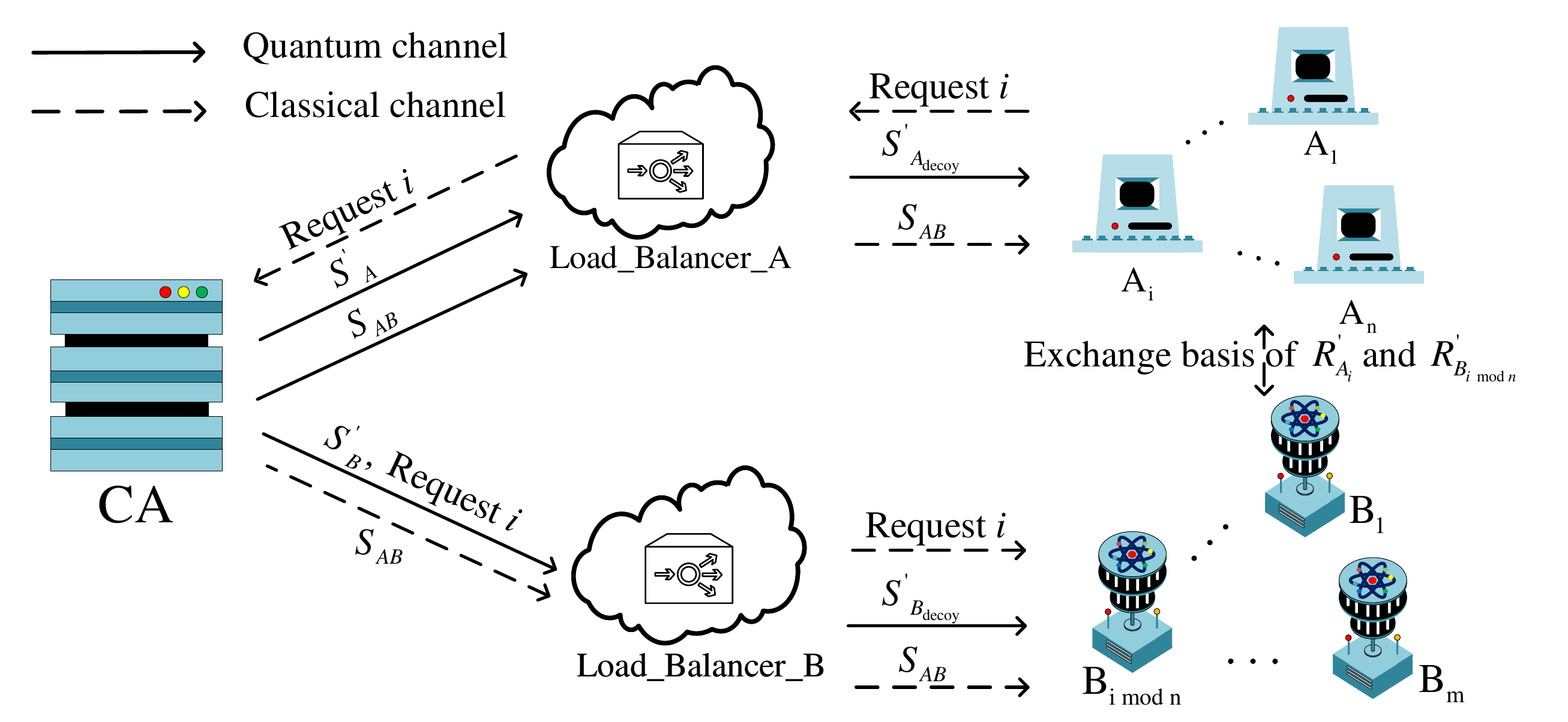}
	\caption{(Color Online) Information flow of identity authentication phase in the RAM-SMVBQC protocol}
	\label{fig2}
\end{figure*}

A2-4 According to Table \ref{tab:3}, $A_i$ and $B_{i\ mod\ n}$ check independently whether the other side's measurement results are correct and complete the mutual identity authentication similar to S2-5 in the reviewed BQC protocol. The information flow of Phase 2 is shown in Fig. \ref{fig2}.

Phase 3: the blind quantum computation phase

A3-1 After $A_i$ and $B_{i\ mod\ n}$ authenticate each other successfully, $B_{i\ mod\ n}$ should generate a $N(\alpha+\beta+1)$-qubit state $\rho$. The state $\rho$ consists of $\alpha+\beta+1$ registers, and each register stores $N$ qubits, where $\alpha$ and $\beta$ are both real numbers. If $B_{i\ mod\ n}$ is honest, he generates a graph state $\vert G\rangle\equiv(\underset{e\in E}{\bigotimes}CZ_e)\vert+\rangle^{\otimes N}$ in each register, which has been introduced in Sec. \ref{Sec2}.

A3-2 To reduce the workload of $B_{i\ mod\ n}$ and improve the efficiency of the protocol, $B_{i\ mod\ n}$ can request for help from other idle servers through Load\_Balancer\_B. To be more specific, $B_{i\ mod\ n}$ broadcasts a request to the other idle servers with the help of Load\_Balancer\_B.

A3-3 If $M$ idle servers respond to the request, $B_{i\ mod\ n}$ announces the information of $\vert G\rangle$. Then, each of them needs to generate the state $\vert G\rangle^{\otimes\frac{\alpha+\beta+1}{M}}$. $B_{i\ mod\ n}$ sends each qubit of $\vert G\rangle^{\otimes\frac{\alpha+\beta+1}{M}}$ to $A_i$ one by one. The graph state generated by idle servers are resent by $B_{i\ mod\ n}$ to $A_i$. Note that, $B_{i\ mod\ n}$ can accept $A_{i+1}$'s request after he generates the graph state and he only needs to send a qubit of $\vert G\rangle$ from his quantum memory.

A3-4 $A_i$ chooses $\alpha$ registers uniform randomly and discards them to guarantee that the remaining state is close to an independent and identically distributed sample by using the quantum de Finetti theorem \cite{li2015quantum}. Then she chooses $\beta$ graph states $\vert G\rangle$ at random for stabilizer test. If the stabilizer test passes, she uses the remaining one graph state $\rho_A$ for MBQC.

$Security$ $and$ $verifiability$ $analysis$. We first show the security of registration phase (Phase 1) and the identity authentication phase (Phase 2), which means that no matter whether there is an external or internal attacker, the client's register key $K_{AB}$ cannot be obtained by attackers in Phase 1. Likewise, the attackers also cannot replace the identity of the client in Phase 2. Then it is necessary to show the protocol in the blind quantum computation phase (Phase 3) satisfies blindness, correctness, and verifiability.

In Phase 1, load balancers and CA are semi-honest parties who should follow the protocol honestly, but can try to extract information by themselves without colluding with other entities. First, CA is considered as an insider attacker who wants to obtain the client's register key $K_{AB}$. CA cannot distinguish the basis of the qubits which were sent by load balancers in A1-2 and A1-3 since the non-orthogonal basis is not distinguishable. Therefore, the probability of detecting dishonest behavior of CA is $\frac{1}{2^{m}}$ where $m$ is the length of $S_A$ and $S_B$. For example, if the qubits resent by Load\_Balancer\_A and Load\_Balancer\_B from $A_i$ and $B_{i\ mod\ n}$ are $\vert0\rangle_A$ and $\vert0\rangle_B$ respectively, the measurement result of the CA should be $\vert\psi^{+}\rangle_{AB}$ or $\vert\psi^{-}\rangle_{AB}$ based on Table \ref{tab2}. The malicious CA will return a random result and the probability of returning an incorrect result is $\frac{1}{2}$. Similarly, Load\_Balancer\_A and Load\_Balancer\_B can be regarded as particle generators to randomly prepare single photon sequences and their roles in the quantum network are only to reduce the quantum capability of the clients. If they are dishonest and seen as insider attackers, their malicious behaviors can be detected with a certain probability due to the decoy detection in A1-2 and A1-3. As shown above, an insider cannot obtain the register key $K_{AB}$. In addition, if there exist an external attacker Eve, who can intercept the qubit sequence $S_A$ and $S_B$ from Load\_Balancer\_A and Load\_Balancer\_B and make Bell measurements on them. But based on the measurement results, Eve can only know whether the quantum states of $A_i$ and $B_{i\ mod\ n}$ are the same or opposite and the shared key $K_{AB}$ cannot be deduced. For example, if Eve obtains the result $\vert\psi^{-}\rangle_{AB}$, then he can only get that the quantum states of $A_i$ and $B_{i\ mod\ n}$ are the opposite. Eve cannot judge whether the quantum state is prepared in X basis or Z basis. To sum up, neither insider nor external attackers can obtain the value of the shared key $K_{AB}$ in Phase 1.

In Phase 2, if CA is dishonest and considered as an insider attacker, he sends wrong states to $A_i$ and $B_{i\ mod\ n}$. However, his dishonest behavior will be detected because $A_i$ and $B_{i\ mod\ n}$ need to announce the measurement results and check each other's measurement results independently. Besides, even if Load\_Balancer\_A and Load\_Balancer\_B are regarded as insider attackers, they still cannot know the value of $K_{AB}$ and thus cannot pretend to be $A_i$ or $B_{i\ mod\ n}$. From the view of an outside attacker Eve, he cannot distinguish the intercepted non-orthogonal states that CA sent due to Heisenberg's uncertainty principle. Furthermore, decoy check is used during each transmission. Then, any operation on the quantum state may change the measurement result and the client and the server will discover the attacks with a non-zero probability. Therefore, the proposed protocol can avoid the intercept-resend attack and the entangle-measure attack.

In Phase 3, a single-server BQC method is used, where only a one-way quantum communication from $B_{i\ mod\ n}$ to $A_i$ is needed. The blindness is guaranteed by the no-signaling principle \cite{SanduPopescu1994QUANTUMNA}, which is more fundamental than quantum mechanics. Therefore, the server cannot get anything about the client's input, output and algorithm except the size of the graph state $\vert G\rangle$. Next, we will show that the protocol satisfies $\epsilon$-$verifiability$, namely that the probability that client accepts an incorrect result should be bounded by $\epsilon$. It is easy to see that if $B_{i\ mod\ n}$ behaves honestly, $A_i$ will pass the stabilizer test and accept the outcome of the computation with probability 1. If $A_i$ passes the test, the state $\rho_A$ which can be used for computation satisfies
\begin{equation}
	\begin{aligned}
		\label{eq:4}
		\langle G\vert\rho_A\vert G\rangle \geq 1-\frac{1}{N}=\epsilon,
	\end{aligned}
\end{equation}
with probability at least $1-\frac{1}{N}$ \cite{takeuchi2018verification}. It means that if $A_i$ passes the test, then the state $\rho_A$ used for computation is close to the ideal state $\vert G\rangle$. Otherwise, the computation state $\rho_A$ is far from the ideal state and $A_i$ will reject the computation result. Then a short proof is given for verifiability based on the work in Refs. \cite{takeuchi2018verification,sato2019arbitrable}.

\proof For any $N$-qubit state $\delta$, we can get
\begin{equation}
	\begin{aligned}
		\label{eq:8}
		Tr[(T^{\otimes\beta}\otimes\Pi^{\perp})\delta^{\otimes \beta+1}] \leq \frac{1}{2N^2}.
	\end{aligned}
\end{equation}
Note that $\Pi^{\perp}=I^{\otimes N}-\vert G\rangle\langle G\vert$ and $T$ is a POVM that can make $A_i$ accept the test. Then we can get
\begin{equation}
	\begin{aligned}
		\label{eq:9.1}
		Tr(T\delta)=\frac{1}{2}+\frac{1}{2} \langle G\vert\delta\vert G\rangle,
	\end{aligned}
\end{equation}
\begin{equation}
	\begin{aligned}
		\label{eq:10}
		Tr(\Pi^{\perp}\delta) &=1- \langle G\vert\delta\vert G\rangle \\
		&=2(1-Tr(T\delta)).
	\end{aligned}
\end{equation}
According to Eq. (\ref{eq:9.1}) and Eq. (\ref{eq:10}), we obtain
\begin{equation}
	\begin{aligned}
		\label{eq:11}
		Tr[(T^{\otimes \beta}\otimes\Pi^{\perp})\delta^{\otimes\beta+1}] &= Tr(T\delta^{\beta})Tr(\Pi^{\perp})\\
		&=2Tr(T\delta)^{\beta}(1-Tr(T\delta)).
	\end{aligned}
\end{equation}
Let $Tr(T\delta)=\frac{\beta}{\beta+1}$, Eq. (\ref{eq:11}) can achieve the maximum
\begin{equation}
	\begin{aligned}
		\label{eq:12}
		&2Tr(T\delta)^{\beta}(1-Tr(T\delta))\\ =&	2(\frac{\beta}{\beta+1})^\beta(1-\frac{\beta}{\beta+1})\leq\frac{1}{2N^2}.
	\end{aligned}
\end{equation}
Then we have
\begin{equation}
	\begin{aligned}
		Tr[(T^{\otimes \beta}\otimes\Pi^{\perp})\rho^{\otimes\beta+1}] &= Tr(T^{\otimes \beta}\rho)Tr(\Pi^{\perp}\rho_A)\\ &\leq \frac{1}{2N^2}+\frac{1}{2N^2}=\frac{1}{N^2}.
	\end{aligned}
\end{equation}
If $Tr(\Pi^{\perp}\rho_A) \geq \frac{1}{N}$, then $Tr(T^{\otimes \beta}\rho) \leq \frac{1}{N}$. Therefore
\begin{equation}
	\begin{aligned}
		\label{eq.14}
		Tr(T^{\otimes \beta}\rho)Tr(\Pi^{\perp}\rho_A)=Pr(accept)E(1-\langle G\vert\rho_A\vert G\rangle),
	\end{aligned}
\end{equation}
where $E(\cdot)$ is the expected value and $Pr(accept)$ is the probability that $A_i$ accepts the test. For Eq. (\ref{eq.14}), according to Markov inequality
\begin{equation}
	\begin{aligned}
		\frac{1}{N}Pr[(1-\langle G\vert\rho_A\vert G\rangle) > \frac{1}{N}]\leq E(1-\langle G\vert\rho_A\vert G\rangle) \leq \frac{1}{N},
	\end{aligned}
\end{equation}
it is easy to deduce
\begin{equation}
	\begin{aligned}
		Pr(accept)Pr[(1-\langle G\vert\rho_A\vert G\rangle) > \frac{1}{N}] \leq \frac{1}{N}
	\end{aligned}
\end{equation}
which means $\langle G\vert\rho_A\vert G\rangle \geq 1-\frac{1}{N}$ with probability $1-\frac{1}{N}$ when $A_i$ accepts the test.

\subsection{The second secure multi-party VBQC protocol based on the prepare-and-send method}
\label{Sec4.2}

Next we introduce the second secure multi-party VBQC protocol based on the prepare-and-send method (called PAS-SMVBQC) where the clients can only generate single qubits. In this protocol, the clients also have to register first and then delegate their computational tasks to remote quantum servers according to register keys. The specific steps are given as follows.

Phase 1: the registration phase

B1-1 The client $A_i$ sends a request $i$ to Load\_Balancer\_A who puts it into the request queue. According to the FIFO principle, Load\_Balancer\_A retransmits the request $i$ to Load\_Balancer\_B.

B1-2 The client $A_i$ and the server $B_{i\ mod\ n}$ prepare a series of single quantum states randomly in X basis or Z basis which are denoted by $S_A$ and $S_B$, respectively, and the length of them is $k$. Then they insert decoy states in $\{\vert0\rangle,\vert1\rangle,\vert+\rangle,\vert-\rangle\}$ with fixed length into $S_A$ and $S_B$ and send the new sequences $S_{A_{decoy}}$ and $S_{B_{decoy}}$ to Load\_Balancer\_A and Load\_Balancer\_B. After they receive these qubits, $A_i$ and $B_{i\ mod\ n}$ tell positions and the basis of decoy qubits. Load\_Balancer\_A and Load\_Balancer\_B measure them in the informed measurement bases. If the error rate exceeds the predefined threshold, $A_i$ has to terminate the protocol and move to B1-1.

B1-3 CA performs Bell measurements on $S_A$ and $S_B$, which are sent from Load\_Balancer\_A and Load\_Balancer\_B, and records the measurement results as $R_{AB}$. CA sends $R_{AB}$ to Load\_Balancer\_A and Load\_Balancer\_B across public classical channels. Then Load\_Balancer\_A resends $R_{AB}$ to $A_i$ and Load\_Balancer\_B resends $R_{AB}$ to $B_{i\ mod\ n}$.

B1-4 When $A_i$ and $B_{i\ mod\ n}$ receive $R_{AB}$ successfully, they have to securely exchange the bases of qubits in $S_A$ and $S_B$. Similar to A1-5, they keep the $j$-th bit under the same basis as the initial raw key bit $K_{AB}^j \in \{0,1\}$ when $R_{AB}^j=11$.

B1-5 $A_i$ and $B_{i\ mod\ n}$ have to pick a portion of their raw keys to estimate error rate and detect eavesdropping because of the effect of noise in the actual channel. If the error rate is below the threshold, the channel is secure. Otherwise, $A_i$ has to terminate the protocol and move to B1-1.

B1-6 The steps from B1-1 to B1-5 should be repeated until all clients have completed the registration and each of them has obtained a shared key with a specific network server.

Phase 2: the identity authentication phase

\begin{figure*}[!t]
	\centering
	\includegraphics[width=0.8\textwidth]{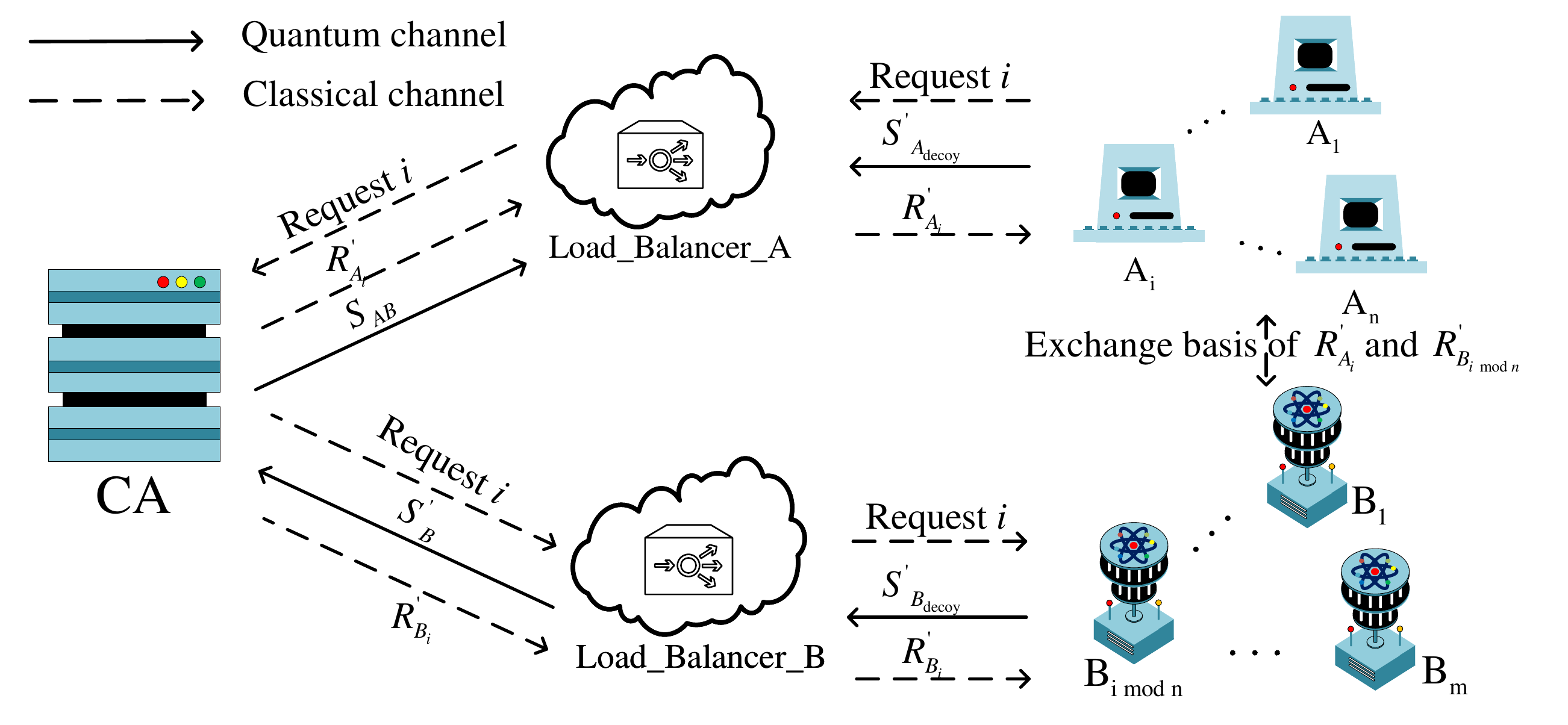}
	\caption{ (Color Online) Information flow of identity authentication phase in the PAS-SMVBQC protocol}
	\label{fig3}
\end{figure*}

B2-1 As shown in Fig. \ref{fig3}, if a registered client $A_i$ wants to delegate a quantum computational task to a remote server, he needs to send a request $i$ to Load\_Balancer\_A firstly.

B2-2 According to the FIFO principle, Load\_Balancer\_A resends the request $i$ to Load\_Balancer\_B and CA, and Load\_Balancer\_B resends the request to $B_{i\ mod\ n}$.

B2-3 $A_i$ and $B_{i\ mod\ n}$ prepare single qubit sequences $S_A^{'}$ and $S_B^{'}$ randomly in X basis or Z basis according to the raw key $K_{AB}$. The encoding method is shown in Table 4.

Then they also prepare decoy sequences in $\{\vert0\rangle,\vert1\rangle,\vert+\rangle,\vert-\rangle\}$ with fixed length and insert into $S_A^{'}$ and $S_B^{'}$ to generate new sequences $S^{'}_{A_{decoy}}$ and $S^{'}_{B_{decoy}}$, which should be sent to Load\_Balancer\_A and Load\_Balancer\_B.

B2-4 When Load\_Balancer\_A and Load\_Balancer\_B receive $S^{'}_{A_{decoy}}$ and $S^{'}_{B_{decoy}}$, they both need to measure the decoy qubits, the positions and basis of which are informed by $A_i$ and $B_{i\ mod\ n}$. If the eavesdropping detection passes, they transmit $S_A^{'}$ and $S_B^{'}$ to CA.

B2-5 After CA receives the sequences $S_A^{'}$ and $S_B^{'}$, he randomly generates $\frac{k_{AB}}{2}$ bits, where $k_{AB}$ is the length of $K_{AB}$ and the $j$-th bit is denoted by $M_j\in\{0,1\}$ for $j\in\{0,1,...,\frac{k_{AB}}{2}\}$. If $M_j=0$, he measures the $j$-th qubit of $S_A^{'}$ in Z basis and records the measurement results $R^{'j}_A$; otherwise he measures it in X basis. He also needs to measure $S_B^{'}$ according to the value of $M$ and records the measurement results as $R^{'}_B$. Then, CA announces the value of $R^{'}_A$ and $R^{'}_B$.

B2-6 According to $R^{'}_A$ and $R^{'}_B$, $A_i$ and $B_{i\ mod\ n}$ could check the identity of the other party. For example, assume that $K_{AB}$ is an $8$-bit string and $K_{AB}=01001011$, as shown in Table 4, the qubit sequence of $S_A^{'}$ and $S_B^{'}$ are $\{\vert1\rangle,\vert0\rangle,\vert+\rangle,\vert-\rangle\}$. If CA generates $M=0101$ and the measurement result $R^{'}_A=1011$, according to Table 4 and the value of $K_{AB}$, $A_i$ can confirm the identity of $B_{i\ mod\ n}$ when $R^{'1}_{B}=1$ and $R^{'4}_{B}=1$.

\begin{table*}[!htb]
	\tabcolsep 6pt
	\centering
	\label{tab3}
	\caption{The Rules of Code $S^{'}_{A}$ and $S^{'}_{B} $ according to $K_{n}K_{n+1}$ and Corresponding Measurement Results $R^{'}_{A}$ and $R^{'}_{B}$ }\vspace{-2mm}
	{\footnotesize
		\begin{tabular*}{\linewidth}{p{60pt}|p{21pt}|p{21pt}|p{21pt}|p{21pt}|p{21pt}|p{21pt}|p{21pt}|p{21pt}}
			\cline{1-9}	
			$K_{n}K_{n+1}$ & \multicolumn{2}{c|}{00}  & \multicolumn{2}{c|}{01} & \multicolumn{2}{c|}{10} & \multicolumn{2}{c}{11}\\
			\cline{2-9}
			$S^{'}_{A}$ and $S^{'}_{B} $ & \multicolumn{2}{c|}{$\vert0\rangle$} & \multicolumn{2}{c|}{$\vert 1\rangle$} &\multicolumn{2}{c|}{$\vert+\rangle$} & \multicolumn{2}{c}{$\vert -\rangle$}\\
			\cline{1-9}
			The value of $M_j$ & 0 & 1 & 0 &  1 & 0 & 1 & 0 & 1 \\\cline{2-9}
			Measurement basis & $Z$ & $X$ & $Z$ & $X$ & $Z$ & $X$ & $Z$ & $X$ \\\cline{2-9}
			$R^{'}_{A}$ and $R^{'}_{B} $ & 0 & 0 or 1 & 1 & 0 or 1 & 0 or 1 & 0 & 0 or 1 & 1
			\\\hline
		\end{tabular*}
	}
\end{table*}

Phase 3: the blind quantum computation phase

B3-1 After $A_i$ and $B_{i\ mod\ n}$ authenticate each other successfully, $A_i$ wants to ask the quantum server $B_{i\ mod\ n}$ to help her perform the task of VBQC.

B3-2 $A_i$ generates $C+T$ qubits and the state of each qubit is $\vert+_{\theta_i}\rangle$=$\frac{1}{\sqrt{2}}(\vert0\rangle+ e^{i\theta_i}\vert1\rangle)(i=1,2,...,N)$, where $\theta_i$ is uniformly and randomly chosen from the set $\{0,\pi/4,2\pi/4,...,7\pi/4\}$. Note that, $C$ is the number of computation qubits used for performing computation and $T$ is the number of trap qubits used for verification. $A_i$ should also generate $D$ dummy qubits which are chosen at random from $\{\vert0\rangle,\vert1\rangle\}$. Then $A_i$ sends these qubits to Bob.

B3-3 $B_{i\ mod\ n}$ generates graph state $\vert G\rangle$ as required by Alice, where $\vert G\rangle$ is an $N$-qubit universal graph state and $N=C+T+D$.

B3-4 $A_i$ asks $B_{i\ mod\ n}$ to measure these qubits. If a qubit in $\vert G\rangle$ is uesd for computation, then it will be measured in the basis $\{\vert0\rangle\pm e^{i\delta_i}\vert1\rangle\} $ where $\delta_i=\phi_i^{'}+\theta_i + r_i\pi$, $\phi_i^{'}$ is obtained according to the previous measurements, and $ r_i$ is chosen from the set $\{0,1\}$ at random. However, trap qubits will be measured in the basis $\{\vert0\rangle\pm e^{i\theta_i+r_i\pi}\vert1\rangle\}$ and dummy qubits are measured in the basis randomly chosen from $\{\vert0\rangle\pm \vert1\rangle,\vert0\rangle\pm e^{i\pi/4}\vert1\rangle, \vert0\rangle\pm e^{i2\pi/4}\vert1\rangle, \vert0\rangle\pm e^{i3\pi/4}\vert1\rangle\}$.

B3-5 If all the measurement results of trap qubits are correct, $A_i$ recovers the outcome of the computation as she knows the value of each random number $r_i$.

$Security$ $and$ $verifiability$ $analysis$. Similar to the RAM-SVMBQC protocol, both CA and load balancers are regarded as semi-honest. The insider attacks and outsider attacks of Phase 1 and Phase 2 will be analyzed. In addition, the correctness, blindness and verifiability of Phase 3 are also discussed in the following.

Clients can only measure single qubits in the first RAM-SVMBQC protocol and they must ask the load balancers for help to generate single qubits. However, in the second PAS-SMVBQC protocol, the clients could generate single qubit by themselves. The role of load balancers is only to resend qubits. If they are dishonest and seen as insider attackers, their malicious behaviors will be detected with a non-zero probability in B1-2 because of the decoy detection. Besides, similar to the RAM-SVMBQC, if there exist an insider attack CA, he cannot distinguish the basis of $S_A$ and $S_B$ and his malicious behaviors will be detected with probability $\frac{1}{2^n}$, where $n$ is the length of $S_A$ and $S_B$. Furthermore, due to decoy detection in each transmission, the proposed protocol can defend against external attackers. In Phase 2, CA only needs to measure $S_A$ and $S_B$ according to the values of $M$, which are generated by himself. Neither a dishonest CA nor an outside attacker Eve can get any information without being detected. Thus, the security of the client and the server can be guaranteed by using the initial security key shared between the client and the server.

In Phase 3, if $B_{i\ mod\ n}$ behaves honestly, then all the measurements on trap qubits will produce correct results and the computation will be performed correctly. Even if $B_{i\ mod\ n}$ behaves maliciously, he cannot get any information about $A_i$ as well. Then, we will prove the blindness of the protocol.

\proof The client's quantum input is blind for server, as the state that $B_{i\ mod\ n}$ received from $A_i$ is
\begin{equation}
	\begin{aligned}
		\label{eq:8.1}
		&\frac{1}{10}\sum_{\theta_i}(\vert+_{\theta_i}\rangle\langle+_{\theta_i}\vert+\vert0\rangle\langle 0\vert+\vert1\rangle\langle 1\vert)\\
		=& \frac{1}{10}(\vert+\rangle\langle +\vert+\cdots+\vert+_{\frac{7\pi}{4}}\rangle\langle +_{\frac{7\pi}{4}}\vert+\vert0\rangle\langle 0\vert+\vert1\rangle\langle 1\vert)\\
		=&\frac{I}{2},
	\end{aligned}
\end{equation}
which is a maximally mixed state.

Then we show the client $A_i$'s algorithm and output are also hidden from the server by using the method similar to that in Refs. \cite{TomoyukiMorimae2015GroundSB,TomoyukiMorimae2012BlindTM}.

\proof Let $\Omega_i=\{\delta_i\}_{i=1}^{N}$ is the set of angles that $A_i$ sends to $B_{i\ mod\ n}$, $\Theta_i=\{\theta_i\}_{i=1}^{N}$ is the set of $A_i$'s privacy, and $R_j=\{ r_i\}_{i=1}^{N}\in\{0,1\}$ is a hidden binary parameters chosen by $A_i$. Let $\{\Pi_i\}_{i=1}^m$ be a POVM which is performed by $B_{i\ mod\ n}$, and $\Lambda\in\{1\dots m\}$ are the corresponding results of the POVM. Suppose that $B_{i\ mod\ n}$ wants to know $A_i$'s privacy. Then we have

\begin{equation}
\begin{aligned}
	\label{eq:8.2}
	&P(\Theta_i=\{\theta_i\}_{i=1}^{N}\vert\Lambda =i, \Omega_i=\{\delta_i\}_{i=1}^{N})\\
	=& \frac{P(\Lambda=i\vert\Theta_i=\{\theta_i\}_{i=1}^{N},\Omega_i=\{\delta_i\}_{i=1}^{N})P(\Theta_i=\{\theta_i\}_{i=1}^{N},\Omega_i=\{\delta_i\}_{i=1}^{N})}{P(\Lambda=i,\Omega_i=\{\delta_i\}_{i=1}^{N}))}\\
	=&\frac{P(\Lambda=i\vert\Theta_i=\{\theta_i\}_{i=1}^{N},\Omega_i=\{\delta_i\}_{i=1}^{N})P(\Theta_i=\{\theta_i\}_{i=1}^{N})P(\Omega_i=\{\delta_i\}_{i=1}^{N})}{P(\Lambda=i\vert\Omega_i=\{\delta_i\}_{i=1}^{N})P(\Omega_i=\{\delta_i\}_{i=1}^{N})}\\
	=&P(\Theta_i=\{\theta_i\}_{i=1}^{N})\frac{P(\Lambda=i\vert\Theta_i=\{\theta_i\}_{i=1}^{N},\Omega_i=\{\delta_i\}_{i=1}^{N})}{P(\Lambda=i\vert\Omega_i=\{\delta_i\}_{i=1}^{N})}\\
	=&P(\Theta_i=\{\theta_i\}_{i=1}^{N}),
\end{aligned}
\end{equation}

which means that $B_{i\ mod\ n}$ cannot know the algorithms of $A_i$. Similarly, the blindness of the client $A_i$'s output can be proved as

\begin{equation}
\begin{aligned}
	\label{eq:8.3}
	&P(R_j=\{ r_i\}_{i=1}^{N}\vert\Lambda=i,\Theta_i=\{\theta_i\}_{i=1}^{N})\\
	=&\frac{P(\Lambda=i\vert R_j=\{ r_i\}_{i=1}^{N},\Omega_i=\{\delta_i\}_{i=1}^{N})P(R_j=\{ r_i\}_{i=1}^{N},\Theta_i=\{\theta_i\}_{i=1}^{N})}{P(\Lambda=i,\Theta_i=\{\theta_i\}_{i=1}^{N})}\\
	=&\frac{P(\Lambda=i\vert R_j=\{ r_i\}_{i=1}^{N}, \Theta_i=\{\theta_i\}_{i=1}^{N} )P(R_j=\{ r_i\}_{i=1}^{N})P(\Theta_i=\{\theta_i\}_{i=1}^{N})}{P(\Lambda=i\vert\Omega_i=\{\delta_i\}_{i=1}^{N})P(\Theta_i=\{\theta_i\}_{i=1}^{N})}\\
	=&P(R_j=\{ r_i\}_{i=1}^{N}) \frac{P(\Lambda=i\vert R_j=\{ r_i\}_{i=1}^{N},\Theta_i=\{\theta_i\}_{i=1}^{N})}{P(\Lambda=i\vert\Theta_i=\{\theta_i\}_{i=1}^{N})}\\
	=&P(R_j=\{ r_i\}_{i=1}^{N}).
\end{aligned}
\end{equation}

In addition, $B_{i\ mod\ n}$ does not know the positions and prepared basis of trap qubits. Thus he cannot determine the measurement result of trap qubits and his malicious behaviors could be found by $A_i$ with a non-zero probability.

We make a brief description about the verifiability similar to that in Ref. \cite{fitzsimons2017unconditionally}. Consider a single trap qubit $(T = 1)$ at a uniformly random position in $\vert G\rangle$, denoted as $\vert+_{\theta_t}\rangle$. Then, we can get that the probability $Pr(incorrect, accept)$ that $A_i$ accepts an incorrect output satisfies the joint distribution

\begin{equation}
	\begin{aligned}
		\label{eq:9}
		Pr(incorrect, accept)\leq \epsilon=1-\frac{1}{2N}
	\end{aligned}
\end{equation}
for quantum output, where $N$ is the number of the graph state $\vert G\rangle$. However, if the output of the protocol is classical, this bound satisfies

\begin{equation}
	\begin{aligned}
		\label{eq:9}
		Pr(incorrect, accept)\leq \epsilon= \frac{N-1}{N}=1-\frac{1}{N}.
	\end{aligned}
\end{equation}
In addition, if $A_i$ should ask $B_{i\ mod\ n}$ to generate a resource state called dotted-triple graph \cite{ElhamKashefi2017OptimisedRC}, the number of trap qubits can be a constant fraction of the total number of qubits, and the bound improves to $\epsilon = 8/9$. If the protocol is repeated $d$ times, it can be shown that $\epsilon = (8/9)^d$.

\subsection{The third multi-party VBQC protocol with the circuit-based model}
\label{Sec4.3}
By using the circuit model similar to that in Ref. \cite{li2021blind}, the third secure multi-party verifiable blind quantum computation protocol (called CB-SMVBQC) is proposed. In this protocol, if the clients can only perform quantum gates $H$ and $\sigma_z^{1/4}$, they can also remotely access a quantum server with the help of load balancer and the CA, and delegate their quantum computational tasks to the server. The specific steps of the proposed protocol are as follows.

Phase 1: the registration phase

C1-1 The client $A_i$ sends a request $i$ to Load\_Balancer\_A who puts it into the request queue. According to the FIFO principle, Load\_Balancer\_A retransmits the request $i$ to Load\_Balancer\_B. Load\_Balancer\_A also prepares $m+k$ states $\vert0\rangle$ which form the sequence $S_{LA}$ and send them to $A_i$.

C1-2 $A_i$ randomly generates $\vert0\rangle=\vert0\rangle$, $\vert1\rangle=H(\sigma_z^{1/4})^4 H\vert0\rangle $, $\vert+\rangle=H\vert0\rangle$, or $\vert-\rangle=(\sigma_z^{1/4})^4H\vert0\rangle$ by applying some quantum gates such as $H$ and $\sigma_z^{1/4}$ on each received qubit of $S_{LA}$. Then $A_i$ returns the new qubit sequences $S_{A_{decoy}}$ to Load\_Balancer\_A. After Load\_Balancer\_A receives these qubits, $A_i$ tells the positions and bases of $m$ qubits. Load\_Balancer\_A measures them in the informed measurement basis and the remaining qubits are expressed as $S_{A}$ which should be sent to CA. If the error rate is higher than the predefined threshold, an eavesdropper Eve is considered to exist in the channel and the protocol should restart.

C1-3 $B_{i\ mod\ n}$ randomly prepares $m+k$ single qubits in $\{\vert0\rangle,\vert1\rangle,\vert+\rangle,\vert-\rangle\}$, which compose the qubit sequence $S_{B_{decoy}}$ and will be sent to Load\_Balancer\_B. After eavesdropping detection, the remaining qubits $S_{B}$ should be sent to CA. Similar to that in C1-2, if the error rate is higher than the predefined threshold, eavesdropping may exist and the protocol need start again.

C1-4 CA performs Bell measurements on the qubit sequences $S_A$ and $S_B$ and records the measurement results $R_{AB}$ $\in$ $\{\vert\phi^+\rangle\rightarrow00,\vert\psi^+\rangle\rightarrow01,\vert\phi^-\rangle\rightarrow10,\vert\psi^-\rangle\rightarrow11\}$. CA sends $R_{AB}$ to Load\_Balancer\_A and Load\_Balancer\_B across public classical channels. Then they resend $R_{AB}$ to $A_i$ and $B_{i\ mod\ n}$.

C1-5 When $A_i$ and $B_{i\ mod\ n}$ receive $R_{AB}$, they have to securely exchange the bases of qubits in $S_A$ and $S_B$. Similarly, according to Table 2, they keep the $j$-th bit under the same basis as the initial raw key bit $K_{AB}^j \in \{0,1\}$ when $R_{AB}^j=11$.

C1-6 $A_i$ and $B_{i\ mod\ n}$ need to pick a portion of their raw key bits to estimate error rate and detect eavesdropping because of the effect of noise in the actual channel. If the error rate exceeds the threshold, $A_i$ has to terminate the protocol and move to C1-1 again.

C1-7 The steps from C1-1 to C1-6 should be repeated until all clients have completed the registration and each of them has shared a key with a specific network server. Note that, if $m>n$ and $i>n$, $B_{i\ mod\ n}$ must store multiple keys in his memories securely.

Phase 2: the identity authentication phase

C2-1 If a registered client $A_i$ wants to securely delegate a quantum computational task to a remote server, he needs to send a request $i$ to Load\_Balancer\_A firstly.

C2-2 According to the FIFO principle, Load\_Balancer\_A resends the request $i$ to Load\_Balancer\_B and CA and Load\_Balancer\_B resends the request to $B_{i\ mod\ n}$.

C2-3 Load\_Balancer\_A prepares $m+k$ qubits in the form of $\vert0\rangle$ and sends them to $A_i$. When each qubit arrives, $A_i$ performs the corresponding single-qubit gates on it to make its state be one of $\{\vert0\rangle,\vert1\rangle,\vert+\rangle,\vert-\rangle\}$, where $\vert0\rangle=\vert0\rangle$, $\vert1\rangle=H(\sigma_z^{1/4})^4 H\vert0\rangle $, $\vert+\rangle=H\vert0\rangle$, and $\vert-\rangle=(\sigma_z^{1/4})^4H\vert0\rangle$. Then $A_i$ returns these qubits which form the sequence $S^{'}_{A_{decoy}}$ to Load\_Balancer\_A.

C2-4 $B_{i\ mod\ n}$ prepares single qubit sequences $S_A^{'}$ and $S_B^{'}$ randomly in X basis or Z basis according to the raw key $K_{AB}$. Then they also prepare decoy sequences in $\{\vert0\rangle,\vert1\rangle,\vert+\rangle,\vert-\rangle\}$ with fixed length and insert into $S_B^{'}$ to make a new sequences $S^{'}_{B_{decoy}}$ which will be transmitted to Load\_Balancer\_B.

C2-5 After Load\_Balancer\_A and Load\_Balancer\_B receive $S^{'}_{A_{decoy}}$ and $S^{'}_{B_{decoy}}$, they both need to measure the decoy qubits, the positions and basis of which are informed by $A_i$ and $B_{i\ mod\ n}$. If the eavesdropping detection passes, they transmit $S_A^{'}$ and $S_B^{'}$ to CA.

C2-6 When CA receives the sequences $S_A^{'}$ and $S_B^{'}$, he randomly generates $\frac{k_{AB}}{2}$ bits, where $k_{AB}$ is the length of $K_{AB}$ and the $j$-th bit is denoted by $M_j\in\{0,1\}$ for $j\in\{0,1,...,\frac{k_{AB}}{2}\}$. If $M_j=0$, he measures the $j$-th qubit of $S_A^{'}$ in Z basis and records the measurement results $R^{'j}_A$, otherwise he measures it in X basis. Similarly, he also needs to measure $S_B^{'}$ and records the measurement results as $R^{'}_B$. Then, CA announces the value of $R^{'}_A$ and $R^{'}_B$.

C2-7 Similar to B2-6, $A_i$ and $B_{i\ mod\ n}$ could check the identity of the other party based on the value of $R^{'}_A$ and $R^{'}_B$.

Phase 3: the blind quantum computation phase

C3-1 After $A_i$ and $B_{i\ mod\ n}$ authenticate each other successfully, she asks the specific server $B_{i\ mod\ n}$ to initiate the task of BQC.

C3-2 $B_{i\ mod\ n}$ sends $N$ qubits in states $\vert0\rangle$ to $A_i$, where $N=C+T+D$. $A_i$ randomly chooses $C$ qubits to be used for computation, $T$ qubits to be considered as trap qubits, and $D$ qubits to be taken as dummy qubits. For computation and trap qubits, she performs the $H$ gate on these qubits and then randomly applies the gate $\sigma_z^{1/4}$ on it for $n$ times in order to make the state of the qubit be $\frac{1}{\sqrt{2}}(\vert0\rangle + e^{\frac{n\pi}{4}i}\vert1\rangle)$, where $ n $ is uniformly chosen from $\{0,1,...,7\}$. For dummy qubits, $A_i$ performs the corresponding single-qubit gates to make them become $\vert0\rangle=\vert0\rangle$ or $\vert1\rangle=H(\sigma_z^{1/4})^4$. All of these qubits will be sent to $B_{i\ mod\ n}$ after $A_i$ generates her expected states.

C3-3 $B_{i\ mod\ n}$ generates an $N$-qubit universal graph state $\vert G\rangle$ as required by $A_i$. Then $A_i$ asks $B_{i\ mod\ n}$ to measure the qubits as follows: computation qubits will be measured in the basis $\{\vert0\rangle\pm e^{i\delta_i}\vert1\rangle\} $ where $\delta_i=\phi_i^{'}+\theta_i + r_i\pi$ and $\phi_i^{'}$ is obtained according to the previous measurements and $ r_i$ is chosen from the set $\{0,1\}$ at random; trap qubits will be measured in the basis $\{\vert0\rangle\pm e^{i\theta_i+r_i\pi}\vert1\rangle\}$; and dummy qubits are measured randomly in the basis chosen from $\{\vert0\rangle\pm \vert1\rangle,\vert0\rangle\pm e^{i\pi/4}\vert1\rangle, \vert0\rangle\pm e^{i2\pi/4}\vert1\rangle, \vert0\rangle\pm e^{i3\pi/4}\vert1\rangle\}$.

C3-4 After all the measurements have been performed, $A_i$ recovers the outcome of the computation by using $r_i$. If all trap measurements succeed, $A_i$ accepts the result; otherwise she rejects.

$Security$ $and$ $verifiability$ $analysis$. Similar to the proposed RAM-SMVBQC and PAS-SMVBQC protocols, Phase 1 and Phase 2 of the CB-SMVBQC protocol can be shown to be secure against the insider attack and outside attack, even if load balancers and CA are both semi-honest. In Phase 3, the proposed CB-SMVBQC also satisfies blindness, correctness and verifiability.

In Phase 1, same as the previous two protocols, even if CA is dishonest, his malicious behavior will be detected with probability $\frac{1}{2^n}$, where $n$ is the length of $S_A$ and $S_B$. In addition, load balancers can be regarded as particle generators to prepare state $\vert0\rangle$. They need to generate initial quantum states and measure the qubits returned by $A_i$. Since the qubits should be measured by load balancers in a random basis $\{\vert0\rangle,\vert1\rangle\}$ or $\{\vert+\rangle,\vert-\rangle\}$, any malicious behavior of load balancers will be discovered with a nonzero probability. Furthermore, in Phase 2, the role of CA is only to measure states sent by Load\_Balancer\_A and Load\_Balancer\_B. The client and server can use the initial shared key $K_{AB}$ to perform mutual identity authentication at the same time. No matter whether there is dishonest CA or Eve in Phase 2, the security of the client and the server can be guaranteed by using the initial security key shared between the client and the server. In Phase 3, if $B_{i\ mod\ n}$ is honest, $A_i$ can get the correct result of her computation. Similarly, according to Eqs. (\ref{eq:8.1}-\ref{eq:8.3}), the blindness of the protocol is guaranteed. Besides, even if $B_{i\ mod\ n}$ is dishonest, he cannot get any useful information about $A_i$. Finally, the verifiability of the proposed protocol is based on trap verfication \cite{fitzsimons2017unconditionally} and $A_i$ could verify her result through embedding trap and dummy qubits in computation.

\section{Comparisons}
\label{Sec5}

We compare three proposed multi-party VBQC protocols in quantum networks that depend on different quantum capabilities of clients with Shan et al.'s multi-party BQC protocol \cite{shan2021multi} and other similar BQC protocols \cite{broadbent2009universal,morimae2013blind} mainly from the trustworthiness of CA, load balancers, and servers, the quantum capability that a client needs, the quantum capability of servers, the verifiability of the computation, and the security of key for identity authentication key as shown in Table \ref{tab4}. Both in the BFK protocol and the MF protocol, only an honest remote server is considered. However, in Shan et al.'s protocol and three proposed protocols, there are multiple parties in quantum networks. But in Shan et al.'s protocol, both the servers and load balancers should be honest, while they can be semi-honest in the three proposed protocols. Besides, only one type of the clients that have the ability to make measurements are considered and they cannot verify their computation results in Shan et al.'s protocol \cite{shan2021multi}. In the proposed protocols, clients' quantum capabilities are flexible and three types of clients with different quantum capabilities have been taken into consideration. Furthermore, clients can both verify the correctness of the computation and the honesty of load balancers and the servers. In addition, in the BFK protocol and the MF protocol, identity authentication between the client and the server is not considered, while in the proposed protocols and Shan et al.'s protocol, mutual authentication among multiple clients and servers is realized. Especially, the raw key shared by clients and the server cannot be obtained by load balancers in the proposed protocols, while it is possible in Shan et al.'s protocol \cite{shan2021multi}.

\begin{table*}[!htb]
	\tabcolsep 2pt
	\centering
	\caption{\label{tab4}Comparisons Between The Proposed Protocols and Other BQC Protocols}
	{\footnotesize
		\begin{tabular*}{\linewidth}{p{50pt}p{55pt}p{50pt}p{50pt}p{50pt}p{50pt}}\hline\hline
			& Trustworthiness of CA, load balancers and servers & The quantum capability of client & The quantum capability of server & Verifiability &  Raw key for identity authentication  \\\hline
			
			Shan et al.'s protocol \cite{shan2021multi}  & Honest load balancers and servers, a semi-honest CA  & Measuring single-qubit states  & Full quantum  & None & Obtained by load balancer, client and server \\
			
			BFK protocol \cite{broadbent2009universal} &  A trust server & Preparing single-qubit states & Full quantum & Trap verification & N/A \\
			
			MF protocol \cite{morimae2013blind} & A trust server & Measuring single-qubit states & Full quantum & None & N/A \\
			
			RAM-SMVBQC protocol &  Semi-honest load balancers, servers and a CA & Measuring single-qubit states  & Full quantum & Stabilizer test & Only obtained by client and server\\
			
			PAS-SMVBQC protocol & Semi-honest load balancers, servers and a CA & Preparing single-qubit states & Full quantum & Trap verification & Only obtained by client and server\\
			
			CB-SMVBQC protocol & Semi-honest load balancers, servers and a CA & Implementing $H$ and $\sigma_z^{1/4}$ gates & Full quantum & Trap verification & Only obtained by client and server
			\\\hline\hline
		\end{tabular*}
	}
\end{table*}

\section{Conclusion}
\label{Sec6}
Different models are suitable for the clients with various quantum capabilities. In order to provide secure delegation of quantum computing for three types of clients with different quantum capabilities in quantum networks, three VBQC protocols with identity authentication have been put forwarded in this paper. In the first protocol, the clients only have to measure a single qubit. In the second protocol, the clients have the ability to generate single qubits. In the third protocol, the clients only have to perform a few single-qubit gates. Besides, in the all three proposed protocols, the CA, servers, and load balancers are considered to be semi-honest, and the load balancers are unable to obtain the shared key between the client and the server. The identity authentication between the clients and the specific servers can also be achieved. Furthermore, the clients with various quantum abilities can delegate their quantum computation in a secure way while they can also verify the correctness of their calculation. However, the consumed quantum resources are still high in the proposed protocols. It deserves further study to improve the efficiency since quantum resources are still precious in quantum networks at present.
\backmatter



\begin{thebibliography}{46}
	\ifx \bisbn   \undefined \def \bisbn  #1{ISBN #1}\fi
	\ifx \binits  \undefined \def \binits#1{#1}\fi
	\ifx \bauthor  \undefined \def \bauthor#1{#1}\fi
	\ifx \batitle  \undefined \def \batitle#1{#1}\fi
	\ifx \bjtitle  \undefined \def \bjtitle#1{#1}\fi
	\ifx \bvolume  \undefined \def \bvolume#1{\textbf{#1}}\fi
	\ifx \byear  \undefined \def \byear#1{#1}\fi
	\ifx \bissue  \undefined \def \bissue#1{#1}\fi
	\ifx \bfpage  \undefined \def \bfpage#1{#1}\fi
	\ifx \blpage  \undefined \def \blpage #1{#1}\fi
	\ifx \burl  \undefined \def \burl#1{\textsf{#1}}\fi
	\ifx \doiurl  \undefined \def \doiurl#1{\url{https://doi.org/#1}}\fi
	\ifx \betal  \undefined \def \betal{\textit{et al.}}\fi
	\ifx \binstitute  \undefined \def \binstitute#1{#1}\fi
	\ifx \binstitutionaled  \undefined \def \binstitutionaled#1{#1}\fi
	\ifx \bctitle  \undefined \def \bctitle#1{#1}\fi
	\ifx \beditor  \undefined \def \beditor#1{#1}\fi
	\ifx \bpublisher  \undefined \def \bpublisher#1{#1}\fi
	\ifx \bbtitle  \undefined \def \bbtitle#1{#1}\fi
	\ifx \bedition  \undefined \def \bedition#1{#1}\fi
	\ifx \bseriesno  \undefined \def \bseriesno#1{#1}\fi
	\ifx \blocation  \undefined \def \blocation#1{#1}\fi
	\ifx \bsertitle  \undefined \def \bsertitle#1{#1}\fi
	\ifx \bsnm \undefined \def \bsnm#1{#1}\fi
	\ifx \bsuffix \undefined \def \bsuffix#1{#1}\fi
	\ifx \bparticle \undefined \def \bparticle#1{#1}\fi
	\ifx \barticle \undefined \def \barticle#1{#1}\fi
	\bibcommenthead
	\ifx \bconfdate \undefined \def \bconfdate #1{#1}\fi
	\ifx \botherref \undefined \def \botherref #1{#1}\fi
	\ifx \url \undefined \def \url#1{\textsf{#1}}\fi
	\ifx \bchapter \undefined \def \bchapter#1{#1}\fi
	\ifx \bbook \undefined \def \bbook#1{#1}\fi
	\ifx \bcomment \undefined \def \bcomment#1{#1}\fi
	\ifx \oauthor \undefined \def \oauthor#1{#1}\fi
	\ifx \citeauthoryear \undefined \def \citeauthoryear#1{#1}\fi
	\ifx \endbibitem  \undefined \def \endbibitem {}\fi
	\ifx \bconflocation  \undefined \def \bconflocation#1{#1}\fi
	\ifx \arxivurl  \undefined \def \arxivurl#1{\textsf{#1}}\fi
	\csname PreBibitemsHook\endcsname
	
	\bibitem{shor1994algorithms}
	\begin{bchapter}
		\bauthor{\bsnm{Shor}, \binits{P.W.}}:
		\bctitle{Algorithms for quantum computation: discrete logarithms and
			factoring}.
		In: \bbtitle{Proceedings of the 35th Annual Symposium on Foundations of
			Computer Science},
		pp. \bfpage{124}--\blpage{134}
		(\byear{1994})
	\end{bchapter}
	\endbibitem
	
	\bibitem{leibfried2003quantum}
	\begin{barticle}
		\bauthor{\bsnm{Leibfried}, \binits{D.}},
		\bauthor{\bsnm{Blatt}, \binits{R.}},
		\bauthor{\bsnm{Monroe}, \binits{C.}}, \betal:
		\batitle{Quantum dynamics of single trapped ions}.
		\bjtitle{Rev Mod Phys}
		\bvolume{75}(\bissue{1}),
		\bfpage{281}--\blpage{324}
		(\byear{2003})
	\end{barticle}
	\endbibitem
	
	\bibitem{blatt2012quantum}
	\begin{barticle}
		\bauthor{\bsnm{Blatt}, \binits{R.}},
		\bauthor{\bsnm{Roos}, \binits{C.F.}}:
		\batitle{Quantum simulations with trapped ions}.
		\bjtitle{Nat Phys}
		\bvolume{8}(\bissue{4}),
		\bfpage{277}--\blpage{284}
		(\byear{2012})
	\end{barticle}
	\endbibitem
	
	\bibitem{krantz2019a}
	\begin{barticle}
		\bauthor{\bsnm{Krantz}, \binits{P.}},
		\bauthor{\bsnm{Kjaergaard}, \binits{M.}},
		\bauthor{\bsnm{Yan}, \binits{F.}}, \betal:
		\batitle{A quantum engineer's guide to superconducting qubits}.
		\bjtitle{Appl Phys Rev}
		\bvolume{6}(\bissue{2}),
		\bfpage{21318}
		(\byear{2019})
	\end{barticle}
	\endbibitem
	
	\bibitem{kjaergaard2020superconducting}
	\begin{barticle}
		\bauthor{\bsnm{Kjaergaard}, \binits{M.}},
		\bauthor{\bsnm{Schwartz}, \binits{M.E.}},
		\bauthor{\bsnm{Braumüller}, \binits{J.}}, \betal:
		\batitle{Superconducting qubits: current state of play}.
		\bjtitle{Annu Rev Condens Matter Phys}
		\bvolume{11}(\bissue{1}),
		\bfpage{369}--\blpage{395}
		(\byear{2020})
	\end{barticle}
	\endbibitem
	
	\bibitem{XiLinWang201818QubitEW}
	\begin{barticle}
		\bauthor{\bsnm{Wang}, \binits{X.L.}},
		\bauthor{\bsnm{Luo}, \binits{Y.H.}},
		\bauthor{\bsnm{Huang}, \binits{H.L.}}, \betal:
		\batitle{18-qubit entanglement with six photons' three degrees of freedom.}
		\bjtitle{Phys Rev Lett}
		\bvolume{120}(\bissue{26}),
		\bfpage{260502}
		(\byear{2018})
	\end{barticle}
	\endbibitem
	
	\bibitem{wang2019boson}
	\begin{barticle}
		\bauthor{\bsnm{Wang}, \binits{H.}},
		\bauthor{\bsnm{Qin}, \binits{J.}},
		\bauthor{\bsnm{Ding}, \binits{X.}}, \betal:
		\batitle{Boson sampling with 20 input photons and a 60-mode interferometer in a
			$10^{14}$-dimensional hilbert space.}
		\bjtitle{Phys Rev Lett}
		\bvolume{123}(\bissue{25}),
		\bfpage{250503}
		(\byear{2019})
	\end{barticle}
	\endbibitem
	
	\bibitem{YuHe2019ATG}
	\begin{barticle}
		\bauthor{\bsnm{He}, \binits{Y.}},
		\bauthor{\bsnm{Gorman}, \binits{S.K.}},
		\bauthor{\bsnm{Keith}, \binits{D.}}, \betal:
		\batitle{A two-qubit gate between phosphorus donor electrons in silicon}.
		\bjtitle{Nature}
		\bvolume{571}(\bissue{7765}),
		\bfpage{371}--\blpage{375}
		(\byear{2019})
	\end{barticle}
	\endbibitem
	
	\bibitem{hensen2020a}
	\begin{barticle}
		\bauthor{\bsnm{Hensen}, \binits{B.}},
		\bauthor{\bsnm{Huang}, \binits{W.}},
		\bauthor{\bsnm{Yang}, \binits{C.H.}}, \betal:
		\batitle{A silicon quantum-dot-coupled nuclear spin qubit}.
		\bjtitle{Nat Nanotechnol}
		\bvolume{15}(\bissue{1}),
		\bfpage{13}--\blpage{17}
		(\byear{2020})
	\end{barticle}
	\endbibitem
	
	\bibitem{childs2005secure}
	\begin{barticle}
		\bauthor{\bsnm{Childs}, \binits{A.M.}}:
		\batitle{Secure assisted quantum computation}.
		\bjtitle{Quantum Inf Comput}
		\bvolume{5}(\bissue{6}),
		\bfpage{456}--\blpage{466}
		(\byear{2005})
	\end{barticle}
	\endbibitem
	
	\bibitem{arrighi2006blind}
	\begin{barticle}
		\bauthor{\bsnm{Arrighi}, \binits{P.}},
		\bauthor{\bsnm{Salvail}, \binits{L.}}:
		\batitle{Blind quantum computation}.
		\bjtitle{Int J Quantum Inf}
		\bvolume{4}(\bissue{5}),
		\bfpage{883}--\blpage{898}
		(\byear{2006})
	\end{barticle}
	\endbibitem
	
	\bibitem{broadbent2009universal}
	\begin{bchapter}
		\bauthor{\bsnm{Broadbent}, \binits{A.}},
		\bauthor{\bsnm{Fitzsimons}, \binits{J.F.}},
		\bauthor{\bsnm{Kashefi}, \binits{E.}}:
		\bctitle{Universal blind quantum computation}.
		In: \bbtitle{Proceeding of the 50th Annual IEEE Symposium on Foundations of
			Computer Science},
		pp. \bfpage{517}--\blpage{526}
		(\byear{2009})
	\end{bchapter}
	\endbibitem
	
	\bibitem{barz2012demonstration}
	\begin{barticle}
		\bauthor{\bsnm{Barz}, \binits{S.}},
		\bauthor{\bsnm{Kashefi}, \binits{E.}},
		\bauthor{\bsnm{Broadbent}, \binits{A.}}, \betal:
		\batitle{Demonstration of blind quantum computing}.
		\bjtitle{Science}
		\bvolume{335}(\bissue{6066}),
		\bfpage{303}--\blpage{308}
		(\byear{2012})
	\end{barticle}
	\endbibitem
	
	\bibitem{morimae2013secure}
	\begin{barticle}
		\bauthor{\bsnm{Morimae}, \binits{T.}},
		\bauthor{\bsnm{Fujii}, \binits{K.}}:
		\batitle{Secure entanglement distillation for double-server blind quantum
			computation}.
		\bjtitle{Phys Rev Lett}
		\bvolume{111}(\bissue{2}),
		\bfpage{20502}
		(\byear{2013})
	\end{barticle}
	\endbibitem
	
	\bibitem{sheng2015deterministic}
	\begin{barticle}
		\bauthor{\bsnm{Sheng}, \binits{Y.B.}},
		\bauthor{\bsnm{Zhou}, \binits{L.}}:
		\batitle{Deterministic entanglement distillation for secure double-server blind
			quantum computation}.
		\bjtitle{Sci Rep}
		\bvolume{5}(\bissue{1}),
		\bfpage{7815}
		(\byear{2015})
	\end{barticle}
	\endbibitem
	
	\bibitem{li2014triple}
	\begin{barticle}
		\bauthor{\bsnm{Li}, \binits{Q.}},
		\bauthor{\bsnm{Chan}, \binits{W.H.}},
		\bauthor{\bsnm{Wu}, \binits{C.}}, \betal:
		\batitle{Triple-server blind quantum computation using entanglement swapping}.
		\bjtitle{Phys Rev A}
		\bvolume{89}(\bissue{4}),
		\bfpage{40302}
		(\byear{2014})
	\end{barticle}
	\endbibitem
	
	\bibitem{kong2016multiple}
	\begin{barticle}
		\bauthor{\bsnm{Kong}, \binits{X.}},
		\bauthor{\bsnm{Li}, \binits{Q.}},
		\bauthor{\bsnm{Wu}, \binits{C.}}, \betal:
		\batitle{Multiple-server flexible blind quantum computation in networks}.
		\bjtitle{Int J Theor Phys}
		\bvolume{55}(\bissue{6}),
		\bfpage{3001}--\blpage{3007}
		(\byear{2016})
	\end{barticle}
	\endbibitem
	
	\bibitem{morimae2013blind}
	\begin{barticle}
		\bauthor{\bsnm{Morimae}, \binits{T.}},
		\bauthor{\bsnm{Fujii}, \binits{K.}}:
		\batitle{Blind quantum computation protocol in which {A}lice only makes
			measurements}.
		\bjtitle{Phys Rev A}
		\bvolume{87}(\bissue{5}),
		\bfpage{50301}
		(\byear{2013})
	\end{barticle}
	\endbibitem
	
	\bibitem{greganti2016demonstration}
	\begin{barticle}
		\bauthor{\bsnm{Greganti}, \binits{C.}},
		\bauthor{\bsnm{Roehsner}, \binits{M.C.}},
		\bauthor{\bsnm{Barz}, \binits{S.}}, \betal:
		\batitle{Demonstration of measurement-only blind quantum computing}.
		\bjtitle{New J Phys}
		\bvolume{18}(\bissue{1}),
		\bfpage{13020}
		(\byear{2016})
	\end{barticle}
	\endbibitem
	
	\bibitem{reichardt2013classical}
	\begin{barticle}
		\bauthor{\bsnm{Reichardt}, \binits{B.W.}},
		\bauthor{\bsnm{Unger}, \binits{F.}},
		\bauthor{\bsnm{Vazirani}, \binits{U.V.}}:
		\batitle{Classical command of quantum systems}.
		\bjtitle{Nature}
		\bvolume{496}(\bissue{7446}),
		\bfpage{456}--\blpage{460}
		(\byear{2013})
	\end{barticle}
	\endbibitem
	
	\bibitem{huang2017experimental}
	\begin{barticle}
		\bauthor{\bsnm{Huang}, \binits{H.L.}},
		\bauthor{\bsnm{Zhao}, \binits{Q.}},
		\bauthor{\bsnm{Ma}, \binits{X.}}, \betal:
		\batitle{Experimental blind quantum computing for a classical client.}
		\bjtitle{Phys Rev Lett}
		\bvolume{119}(\bissue{5}),
		\bfpage{50503}
		(\byear{2017})
	\end{barticle}
	\endbibitem
	
	\bibitem{2015Iterated}
	\begin{barticle}
		\bauthor{\bsnm{Perez-Delgado}, \binits{C.A.}},
		\bauthor{\bsnm{Fitzsimons}, \binits{J.F.}}:
		\batitle{Iterated gate teleportation and blind quantum computation}.
		\bjtitle{Phys Rev Lett}
		\bvolume{114}(\bissue{22}),
		\bfpage{220502}
		(\byear{2015})
	\end{barticle}
	\endbibitem
	
	\bibitem{xu2022universal}
	\begin{barticle}
		\bauthor{\bsnm{Xu}, \binits{H.R.}},
		\bauthor{\bsnm{Wang}, \binits{B.H.}}:
		\batitle{Universal single-server blind quantum computation for classical
			clients}.
		\bjtitle{Laser Phys Lett}
		\bvolume{19}(\bissue{1}),
		\bfpage{15202}
		(\byear{2022})
	\end{barticle}
	\endbibitem
	
	\bibitem{li2021blind}
	\begin{barticle}
		\bauthor{\bsnm{Li}, \binits{Q.}},
		\bauthor{\bsnm{Liu}, \binits{C.}},
		\bauthor{\bsnm{Peng}, \binits{Y.}}:
		\batitle{Blind quantum computation where a user only performs single-qubit
			gates}.
		\bjtitle{Opt Laser Technol}
		\bvolume{142},
		\bfpage{107190}
		(\byear{2021})
	\end{barticle}
	\endbibitem
	
	\bibitem{li2021quantum}
	\begin{barticle}
		\bauthor{\bsnm{Li}, \binits{W.}},
		\bauthor{\bsnm{Lu}, \binits{S.}},
		\bauthor{\bsnm{Deng}, \binits{D.L.}}:
		\batitle{Quantum federated learning through blind quantum computing}.
		\bjtitle{Sci China Phys Mech}
		\bvolume{64}(\bissue{10}),
		\bfpage{100312}
		(\byear{2021})
	\end{barticle}
	\endbibitem
	
	\bibitem{fitzsimons2017unconditionally}
	\begin{barticle}
		\bauthor{\bsnm{Fitzsimons}, \binits{J.F.}},
		\bauthor{\bsnm{Kashefi}, \binits{E.}}:
		\batitle{Unconditionally verifiable blind quantum computation}.
		\bjtitle{Phys Rev A}
		\bvolume{96}(\bissue{1}),
		\bfpage{12303}
		(\byear{2017})
	\end{barticle}
	\endbibitem
	
	\bibitem{fitzsimons2018post}
	\begin{barticle}
		\bauthor{\bsnm{Fitzsimons}, \binits{J.F.}},
		\bauthor{\bsnm{Hajdušek}, \binits{M.}},
		\bauthor{\bsnm{Morimae}, \binits{T.}}:
		\batitle{Post hoc verification of quantum computation}.
		\bjtitle{Phys Rev Lett}
		\bvolume{120}(\bissue{4}),
		\bfpage{40501}
		(\byear{2018})
	\end{barticle}
	\endbibitem
	
	\bibitem{hajdusek2015device}
	\begin{botherref}
		\oauthor{\bsnm{Hajdusek}, \binits{M.}},
		\oauthor{\bsnm{Pérez-Delgado}, \binits{C.A.}},
		\oauthor{\bsnm{Fitzsimons}, \binits{J.F.}}:
		Device-independent verifiable blind quantum computation.
		arXiv:1502.02563
		(2015)
	\end{botherref}
	\endbibitem
	
	\bibitem{morimae2017verification}
	\begin{barticle}
		\bauthor{\bsnm{Morimae}, \binits{T.}},
		\bauthor{\bsnm{Takeuchi}, \binits{Y.}},
		\bauthor{\bsnm{Hayashi}, \binits{M.}}:
		\batitle{Verification of hypergraph states}.
		\bjtitle{Phys Rev A}
		\bvolume{96}(\bissue{6}),
		\bfpage{62321}
		(\byear{2017})
	\end{barticle}
	\endbibitem
	
	\bibitem{takeuchi2018verification}
	\begin{barticle}
		\bauthor{\bsnm{Takeuchi}, \binits{Y.}},
		\bauthor{\bsnm{Morimae}, \binits{T.}}:
		\batitle{Verification of many-qubit states}.
		\bjtitle{Phys Rev X}
		\bvolume{8}(\bissue{2}),
		\bfpage{21060}
		(\byear{2018})
	\end{barticle}
	\endbibitem
	
	\bibitem{hayashi2015verifiable}
	\begin{barticle}
		\bauthor{\bsnm{Hayashi}, \binits{M.}},
		\bauthor{\bsnm{Morimae}, \binits{T.}}:
		\batitle{Verifiable measurement-only blind quantum computing with stabilizer
			testing}.
		\bjtitle{Phys Rev Lett}
		\bvolume{115}(\bissue{22}),
		\bfpage{200502}
		(\byear{2015})
	\end{barticle}
	\endbibitem
	
	\bibitem{preskill2019quantum}
	\begin{barticle}
		\bauthor{\bsnm{Preskill}, \binits{J.}}:
		\batitle{Quantum computing in the {NISQ} era and beyond}.
		\bjtitle{Quantum}
		\bvolume{2},
		\bfpage{79}
		(\byear{2019})
	\end{barticle}
	\endbibitem
	
	\bibitem{zhang2017observation}
	\begin{barticle}
		\bauthor{\bsnm{Zhang}, \binits{J.}},
		\bauthor{\bsnm{Pagano}, \binits{G.}},
		\bauthor{\bsnm{Hess}, \binits{P.}}, \betal:
		\batitle{Observation of a many-body dynamical phase transition with a 53-qubit
			quantum simulator}.
		\bjtitle{Nature}
		\bvolume{551}(\bissue{7682}),
		\bfpage{601}--\blpage{604}
		(\byear{2017})
	\end{barticle}
	\endbibitem
	
	\bibitem{arute2019quantum}
	\begin{barticle}
		\bauthor{\bsnm{Arute}, \binits{F.}},
		\bauthor{\bsnm{Arya}, \binits{K.}},
		\bauthor{\bsnm{Babbush}, \binits{R.}}, \betal:
		\batitle{Quantum supremacy using a programmable superconducting processor}.
		\bjtitle{Nature}
		\bvolume{574}(\bissue{7779}),
		\bfpage{505}--\blpage{510}
		(\byear{2019})
	\end{barticle}
	\endbibitem
	
	\bibitem{GuanYuWang2020EntanglementPF}
	\begin{barticle}
		\bauthor{\bsnm{Wang}, \binits{G.Y.}},
		\bauthor{\bsnm{Long}, \binits{G.L.}}:
		\batitle{Entanglement purification for memory nodes in a quantum network}.
		\bjtitle{Sci China Phys Mech}
		\bvolume{63}(\bissue{2}),
		\bfpage{220311}
		(\byear{2020})
	\end{barticle}
	\endbibitem
	
	\bibitem{ZhiHaoLiu221}
	\begin{barticle}
		\bauthor{\bsnm{Liu}, \binits{Z.H.}},
		\bauthor{\bsnm{Chen}, \binits{H.W.}}:
		\batitle{Universal and general quantum simultaneous secret distribution with
			dense coding by using one-dimensional high-level cluster states}.
		\bjtitle{J Comput Sci Tech}
		\bvolume{36}(\bissue{1}),
		\bfpage{221}
		(\byear{2021})
	\end{barticle}
	\endbibitem
	
	\bibitem{WenJunShi1291}
	\begin{barticle}
		\bauthor{\bsnm{Shi}, \binits{W.J.}},
		\bauthor{\bsnm{Cao}, \binits{Q.X.}},
		\bauthor{\bsnm{Deng}, \binits{Y.X.}}, \betal:
		\batitle{Symbolic reasoning about quantum circuits in {C}oq}.
		\bjtitle{J Comput Sci Tech}
		\bvolume{36}(\bissue{6}),
		\bfpage{1291}
		(\byear{2021})
	\end{barticle}
	\endbibitem
	
	\bibitem{li2018blind}
	\begin{barticle}
		\bauthor{\bsnm{Li}, \binits{Q.}},
		\bauthor{\bsnm{Li}, \binits{Z.}},
		\bauthor{\bsnm{Chan}, \binits{W.H.}}, \betal:
		\batitle{Blind quantum computation with identity authentication}.
		\bjtitle{Phys Lett A}
		\bvolume{382}(\bissue{14}),
		\bfpage{938}--\blpage{941}
		(\byear{2018})
	\end{barticle}
	\endbibitem
	
	\bibitem{shan2021multi}
	\begin{barticle}
		\bauthor{\bsnm{Shan}, \binits{R.T.}},
		\bauthor{\bsnm{Chen}, \binits{X.}},
		\bauthor{\bsnm{Yuan}, \binits{K.G.}}:
		\batitle{Multi-party blind quantum computation protocol with mutual
			authentication in network}.
		\bjtitle{Sci China Inf Sci.}
		\bvolume{64}(\bissue{6}),
		\bfpage{1}--\blpage{14}
		(\byear{2021})
	\end{barticle}
	\endbibitem
	
	\bibitem{barends2014superconducting}
	\begin{barticle}
		\bauthor{\bsnm{Barends}, \binits{R.}},
		\bauthor{\bsnm{Kelly}, \binits{J.}},
		\bauthor{\bsnm{Megrant}, \binits{A.}}, \betal:
		\batitle{Superconducting quantum circuits at the surface code threshold for
			fault tolerance}.
		\bjtitle{Nature}
		\bvolume{508}(\bissue{7497}),
		\bfpage{500}--\blpage{503}
		(\byear{2014})
	\end{barticle}
	\endbibitem
	
	\bibitem{li2015quantum}
	\begin{barticle}
		\bauthor{\bsnm{Li}, \binits{K.}},
		\bauthor{\bsnm{Smith}, \binits{G.}}:
		\batitle{Quantum de {Finetti} theorem under fully-one-way adaptive
			measurements}.
		\bjtitle{Phys Rev Lett}
		\bvolume{114}(\bissue{16}),
		\bfpage{160503}
		(\byear{2015})
	\end{barticle}
	\endbibitem
	
	\bibitem{SanduPopescu1994QUANTUMNA}
	\begin{barticle}
		\bauthor{\bsnm{Popescu}, \binits{S.}},
		\bauthor{\bsnm{Rohrlich}, \binits{D.}}:
		\batitle{Quantum nonlocality as an axiom}.
		\bjtitle{Found Phys}
		\bvolume{24}(\bissue{3}),
		\bfpage{379}--\blpage{385}
		(\byear{1994})
	\end{barticle}
	\endbibitem
	
	\bibitem{sato2019arbitrable}
	\begin{barticle}
		\bauthor{\bsnm{Sato}, \binits{G.}},
		\bauthor{\bsnm{Koshiba}, \binits{T.}},
		\bauthor{\bsnm{Morimae}, \binits{T.}}:
		\batitle{Arbitrable blind quantum computation}.
		\bjtitle{Quantum Inf Process}
		\bvolume{18}(\bissue{12}),
		\bfpage{370}
		(\byear{2019})
	\end{barticle}
	\endbibitem
	
	\bibitem{TomoyukiMorimae2015GroundSB}
	\begin{barticle}
		\bauthor{\bsnm{Morimae}, \binits{T.}},
		\bauthor{\bsnm{Dunjko}, \binits{V.}},
		\bauthor{\bsnm{Kashefi}, \binits{E.}}:
		\batitle{Ground state blind quantum computation on {AKLT} state}.
		\bjtitle{Quantum Inf Comput}
		\bvolume{15}(\bissue{3}),
		\bfpage{200}--\blpage{234}
		(\byear{2015})
	\end{barticle}
	\endbibitem
	
	\bibitem{TomoyukiMorimae2012BlindTM}
	\begin{barticle}
		\bauthor{\bsnm{Morimae}, \binits{T.}},
		\bauthor{\bsnm{Fujii}, \binits{K.}}:
		\batitle{Blind topological measurement-based quantum computation}.
		\bjtitle{Nat Commun}
		\bvolume{3}(\bissue{1}),
		\bfpage{1036}--\blpage{1036}
		(\byear{2012})
	\end{barticle}
	\endbibitem
	
	\bibitem{ElhamKashefi2017OptimisedRC}
	\begin{barticle}
		\bauthor{\bsnm{Kashefi}, \binits{E.}},
		\bauthor{\bsnm{Wallden}, \binits{P.}}:
		\batitle{Optimised resource construction for verifiable quantum computation}.
		\bjtitle{J Phys A-Math Theor}
		\bvolume{50}(\bissue{14}),
		\bfpage{145306}
		(\byear{2017})
	\end{barticle}
	\endbibitem
	
\end{thebibliography}


\end{document}